\documentclass[fleqn,usenatbib]{mnras}

\usepackage{newtxtext,newtxmath}

\usepackage[T1]{fontenc}

\DeclareRobustCommand{\VAN}[3]{#2}
\let\VANthebibliography\thebibliography
\def\thebibliography{\DeclareRobustCommand{\VAN}[3]{##3}\VANthebibliography}
\usepackage{booktabs}
\usepackage[normalem]{ulem}

\usepackage{graphicx}
\usepackage[dvipsnames]{xcolor}	
\usepackage{amsmath}	

\definecolor{darkgreen}{rgb}{0.0, 0.5, 0.0}

\definecolor{chmagenta}{rgb}{0.54, 0.17, 0.88}

\title[Redshift evolution of BBH eccentricity]{Hierarchical Triples vs. Globular Clusters: Binary black hole merger eccentricity distributions compete and evolve with redshift}

\author[Dorozsmai \& Romero-Shaw et al. ]{Andris Dorozsmai,$^{1}$\thanks{E-mail: andras.dorozsmai@nao.ac.jp}
Isobel M. Romero-Shaw,$^{2,3,4}$\thanks{E-mail: isobel.romero-shaw@bristol.ac.uk}
Aditya Vijaykumar,$^{5}$
Silvia Toonen,$^{6}$ 
Fabio Antonini,$^{7}$
\newauthor
Kyle Kremer,$^{8}$
Michael Zevin,$^{9,10,11}$ and
Evgeni Grishin$^{12,13}$
\\
$^{1}$ National Astronomical Observatory of Japan, National Institutes of Natural Sciences, 2-21-1 Osawa, Mitaka, Tokyo 181-8588, Japan\\
$^{2}$ Department of Applied Mathematics and Theoretical Physics, Cambridge CB3 0WA, United Kingdom\\
$^{3}$ Kavli Institute for Cosmology Cambridge, Madingley Road Cambridge CB3 0HA, United Kingdom\\
$^{4}$ H.H. Wills Physics Laboratory, Tyndall Avenue, Bristol BS8 1TL, United Kingdom\\
$^{5}$ Canadian Institute for Theoretical Astrophysics, University of Toronto, 60 St George Street, Toronto, ON M5S 3H8, Canada\\
$^{6}$ Anton Pannekoek Institute for Astronomy, University of Amsterdam, Science Park 904, 1098 XH Amsterdam, Netherlands\\
$^{7}$  Gravity Exploration Institute, School of Physics and Astronomy, Cardiff University, Cardiff , CF24 3AA, United Kingdom \\
$^{8}$ Department of Astronomy and Astrophysics, University of California, San Diego; La Jolla, CA 92093, USA\\
$^{9}$ Adler Planetarium, 1300 South DuSable Lake Shore Drive, Chicago, IL, 60605, USA\\
$^{10}$Center for Interdisciplinary Exploration and Research in Astrophysics (CIERA), Northwestern University, 1800 Sherman Avenue, Evanston, IL, 60201, USA\\
$^{11}$NSF-Simons AI Institute for the Sky (SkAI), 172 East Chestnut Street, Chicago, IL, 60611, USA\\
$^{12}$School of Physics and Astronomy, Monash University, Clayton, VIC 3800, Australia \\
$^{13}$ OzGrav: Australian Research Council Centre of Excellence for Gravitational Wave Discovery, Clayton, VIC 3800, Australia
}

\date{Accepted XXX. Received YYY; in original form ZZZ}

\pubyear{2015}

\begin{document}
\label{firstpage}
\pagerange{\pageref{firstpage}--\pageref{lastpage}}
\maketitle

\begin{abstract}
The formation mechanisms of merging binary black holes (BBHs) observed by the LIGO–Virgo–KAGRA collaboration remain uncertain. Detectable eccentricity provides a powerful diagnostic for distinguishing between different formation channels, but resolving their eccentricity distributions requires the detection of a large number of eccentric mergers. Future gravitational wave detectors such as the Einstein Telescope and Cosmic Explorer will detect tens of thousands of BBH mergers out to redshifts $z \ge 10$, making it critical to understand the redshift-dependent evolution of eccentricity distributions. We simulate this evolution for two key channels: dynamical assembly in globular clusters (GCs), which leads to rapid, eccentric mergers; and hierarchical triples in the field, where three-body dynamics can induce eccentricity in the inner binary. When considering all BBH mergers, the GC channel dominates overall, consistent with previous studies. However, when focusing on mergers with detectable eccentricity in next-generation detectors, we find that hierarchical triples dominate the eccentric merger rate at $0\le z \le 4$, with GC mergers becoming competitive at higher redshifts. Across all model variations, eccentric mergers in the local Universe ($z\lesssim 1$) have significant contributions from field triples, 
 challenging the common view that such systems primarily form in dense environments. We show that, regardless of cluster and stellar evolution uncertainties, hierarchical triples contribute at least 30 per cent of eccentric mergers across a large range of redshifts.

\end{abstract}

\begin{keywords}
gravitational waves -- stars: black holes -- stars: kinematics and dynamics
\end{keywords}



\section{Introduction}
\label{sec:intro}
Measurable orbital eccentricity in a gravitational wave (GW) signal from a binary black hole (BBH) merger 
in the LIGO-Virgo-KAGRA \citep[LVK;][]{Aasi13} frequency range is a robust indication that the binary was driven to merge rapidly due to the gravitational influence of one or more external bodies, in addition to its own GW emission.
Close-to-merger eccentricity can be driven up due to a single external body exerting gravitational influence on the binary, as in the case of stellar triples found in sparsely-populated environments 
\citep[e.g.,][]{SilsbeeTremaine2017, Antonini2017, RodriguezAntonini2018, Fragione2019, Martinez2022, man22,2022MNRAS.516.1406S, vg25}. 
Alternatively, measurable in-band eccentricities can arise due to dynamical interactions of the binary with many external bodies in densely-populated environments, as in the case of globular clusters \citep[GCs; e.g.,][]{Wen2003, Benacquista2013, Antonini2014, Morscher15, Song2016, Abbas2017, Samsing17, Hong2018, Rodriguez2018:formationmassesmergerrates, FragioneKocsis2018, Kremer2019:LisaBand,Choksi2019,  Zevin:binarybinary:2019, AntoniniGieles:2020:cBHBd, Kamlah2022, DallAmico:2023:clusters}, young or open clusters \citep[e.g.][]{Banerjee2010, Banerjee2017:I, DiCarlo2019, DiCarlo2020:impact_of_metallicity, Kumamoto2019, Trani2022, Prieto2022A} or nuclear star clusters 
\citep[NCs; e.g., ][]{AntoniniRasio:2016:GCNSCVesc, Gondan17, Fragione19, DallAmico:2023:clusters}. 
Further external forces can be exerted due to the presence of the gaseous disc of active galactic nuclei \citep[AGN; e.g.,][]{Yang:2019:HierarchicalAGN, McKernan:2020:AGNPredictions, IshibashiGrobner:2020:AGN, Samsing22, Tagawa:2021:massgap, GrishinAGN24, gilbaum25} or by the central supermassive black hole \citep{AntoniniPerets2012, PetrovichAntonini2017, Bao-Minh2018, LiuLaiWang2019, Maeda2023}.
Meanwhile, isolated BBHs that slowly inspiral under only the influence of their own GW emission will lose any trace of detectable orbital eccentricity by the time they enter the LVK frequency range \citep{Peters64, Breivik2016}.

Observations of eccentric BBH mergers enable constraints to be placed on the overall contribution of externally-driven formation channels \citep[e.g.,][]{Zevin:EccentricImplications:2021}. Being able to distinguish the contributions of multiple formation channels to the eccentric merger rate, however, would require at least $\mathcal{O}(100)$ measurably-eccentric BBH detections \citep{Romero-Shaw:FourEccentricMergers:2022}.
Many of the channels that produce eccentric mergers - for example the GC channel - are predicted to yield only a small fraction ($\sim5$ per cent) of detectable mergers with measurable eccentricities in the LVK band,
making this prospect challenging with existing detectors (where the minimum detectable eccentricity for current detectors at a GW signal frequency of 10 hz is about $e_\mathrm{10Hz} \gtrsim 0.05$, see e.g. \citealt{Lower:Eccentricity:2018}).
The active instruments detect $\mathcal{O}(100)$ GW signals per year \citep{GWTC-3}; even with improvements in sensitivity, it would still take $\gtrsim\mathcal{O}(10)$ years to observe $>\mathcal{O}(100)$ measurably-eccentric BBH mergers. 
The upcoming ground-based next-generation (XG) detectors, such as the Einstein Telescope \citep[ET;][]{Maggiore:2020:ET} and Cosmic Explorer \citep[CE;][]{Evans:2021:CosmicExplorer, Evans:2023:CosmicExplorer}, will have greatly improved sensitivities and wider frequency ranges, increasing both the number of expected yearly detections (up to $\mathcal{O}(10^6)$~yr$^{-1}$) and the sensitivity to $e_\mathrm{10Hz}$ \citep[$\gtrsim10^{-3}$]{Lower:Eccentricity:2018, Saini:2024:ETsensitivity}. Thus, with such improvements, it will become possible to distinguish between formation channels that produce eccentric mergers \citep{Romero-Shaw:FourEccentricMergers:2022}. However, such predictions have been made assuming eccentricity distributions that do not vary with redshift. 
This drastically increased detection count expected for XG instruments is the result of their increased observing volume, out to redshift $z=10$ and above for stellar-mass BBHs \citep[e.g.,][]{Maggiore:2020:ET, Evans:2021:CosmicExplorer, Evans:2023:CosmicExplorer}.
It is therefore imperative to investigate how the eccentricity distributions evolve with redshift to establish whether a significant rate of measurably-eccentric GW mergers is still expected at higher redshifts, as well as to assess whether the dominant formation 
channel for detectably-eccentric mergers changes with redshift.

While measurably-eccentric merging BBHs can be confidently attributed to formation channels involving dynamical driving of the orbital properties, further differentiating between evolutionary paths that produce detectably-eccentric GW mergers
is useful. 
If GW sources originating from a specific formation 
channel can be reliably identified, their properties can provide invaluable insights into the progenitors' environments 
\citep[e.g.][]{Romero-Shaw:2021:GCs, FishbachFragione:2023:GCs} or companions \citep{Meiron2017, Romero-Shaw:2023:InferringInterference, Hendriks2024arXiv241108572H, Samsing2024phaseshift}, as well as into the uncertain physical processes that lead to their eventual merger \citep[e.g.][]{Stevenson2019, AntoniniGieles2020}.

The correlation of GW-observable parameters and redshift itself can be a powerful probe of GW source formation \citep{Safarzadeh2019, Callister:2020:ShoutsMurmurs, Romero-Shaw:2021:GCs, Bavera:2022:chieffvsz, Biscoveanu:2022:RedshiftSpins, vanSon:2022:WeightyMatter, FishbachFragione:2023:GCs}. 
The mechanisms that act to produce BBH mergers via different formation routes may vary with redshift in distinct ways.
The properties of the sources they produce, therefore, also have redshift dependence that depends on formation mechanism.
Understanding the redshift dependence of the eccentricity distributions produced by different mechanisms may be key for distinguishing different externally-driven formation channels that produce measurably-eccentric GW sources. 

In this work, we compare the redshift evolution of the eccentricity distributions expected from hierarchical field triples, in which merging BBHs can be driven to high eccentricities due to the gravitational influence of the tertiary object, and from GCs, where mechanisms such as single-single GW capture and three- or four-body interactions can lead to detectably-eccentric BBH mergers.
Specifically, we investigate how the fraction and merger rate of detectably-eccentric GW sources evolve with redshift.
We do this for two different formation channels that produce eccentric mergers, to assess whether XG detectors can constrain the contributions of such formation channels to the total GW merger rate.
We also study variations in dominance for channels producing eccentric GW mergers over redshift, and whether the eccentricity distributions from triples and GCs evolve in observationally distinguishable ways.
We find that the redshift evolution of the eccentric BBH merger rate varies significantly between the two channels, and that in our fiducial model, triples dominate the XG detectably-eccentric merger rate until $z\gtrsim4$.

This paper is structured as follows.
In Section \ref{sec:models}, we detail our modelling of triples and GCs, and describe how the simulation data is converted into a synthetic population representing the merging BBH demographics across a broad redshift range ($0\leq z\leq6$).
We present the redshift evolution of the eccentricity distribution and the eccentric merger rate for both formation channels in Section \ref{sec:ztrend}, explain how this is sensitively entangled with metallicity, show how our results vary with model uncertainties, and discuss the prospects of distinguishing channels that produce eccentric mergers using XG GW detectors. We summarise our conclusions in Section \ref{sec:conclusions}. 

\section{Simulation details}
\label{sec:models}

\subsection{Models}
\subsubsection{Globular clusters}
\label{sec:GC_channel}
In GCs, mass segregation---primarily driven by two-body relaxation \citep[][]{Spitzer1969}---leads to the formation of an extremely dense BH-dominated central region \citep[see, e.g.,][]{Larson1984, SigurdssonHernquist1993, Merritt2004, Chatterjee2013, Wang2016, Kremer2020:role_of_bh_burning}. Within this dense core, bound few-body systems frequently form that engage in strong resonant interactions with each other and surrounding stars \citep[e.g.][]{FregauRasio2007, Samsing:2017:assembly}. These encounters promote the formation of tight binary systems, and subsequently, the formation of GW sources with non-negligible eccentricities at $10$~Hz peak GW frequency \citep[see, e.g.,][]{Benacquista2013}. Numerical simulations carried out in the last decade have identified the following three primary pathways for the formation of merging BBHs within GCs \citep[see, e.g.,][]{SamsingDorazio2018, DorazioSamsing2018, Rodriguez2018:formationmassesmergerrates, Zevin:binarybinary:2019,Kremer2020}:

\textit{Ejected mergers}: When a BBH has binding energy greater than the typical kinetic energy of surrounding BHs and stars, each strong interaction will shrink its orbit and change its eccentricity drawn from a thermal distribution \citep[e.g.][]{Heggie1975, Hut1983}. These interactions also impart a recoil velocity to the binary, proportional to its orbital velocity. As BBHs undergo more interactions, the received recoil velocity increases and can sometimes exceed the escape velocity of the cluster, resulting in the ejection of the binary.  Ejected BBHs with sufficiently short orbital periods and high eccentricities merge due to GWs within the Hubble time. This formation path is expected to produce the lowest-eccentricity BBH mergers among all GC channels, with typical eccentricities at $10$~Hz of $10^{-8} \lesssim e_\mathrm{10Hz} \lesssim 10^{-3}$ \citep[see, e.g.,][]{PortegiesZwart2000, Downing2010, Rodriguez2016, Park2017, Samsing17}. 

\textit{In-cluster mergers}: In certain cases, a BBH in the cluster will reach sufficiently high eccentricity due to a strong interaction, such that it merges due to GWs before the next encounter occurs. We refer to these as \textit{in-cluster} mergers.  These sources have typical eccentricities in the range of $10^{-6} \lesssim e_\mathrm{10Hz} \lesssim 10^{-2}$ \citep[see, e.g.,][]{Samsing17, Rodriguez2018:formationmassesmergerrates, Rodriguez2018:repeatedmergers, Zevin:binarybinary:2019}. 

\textit{GW captures}: Some BBHs develop such high eccentricities \textit{during} the resonant encounter that they merge almost instantaneously due to GW emission near the pericentre. These short-lived binaries lead to the highest-eccentricity BBH mergers in GCs, i.e. $10^{-3}\lesssim e_\mathrm{10Hz} \lesssim 1$ \citep[e.g.,][]{Gultekin2006, Samsing2014, Samsing:2017:assembly, Rodriguez2018:formationmassesmergerrates, Zevin:binarybinary:2019}. In sufficiently compact GCs, GW capture can also occur due to single-single encounters \citep[see e.g.][]{Samsing2020}. Furthermore, merging BBHs in GCs can also form through secular three-body interactions mediated by a tertiary object \citep[][]{Wen2003, Antonini2014} or by the cluster potential \citep[][]{Hamilton2018}.

To model the population arising from evolution inside GCs, we utilize the \texttt{Cluster Monte Carlo (CMC) Catalog} \citep{Kremer2020}. 
This catalogue comprises $148$ cluster models, of which we use $144$,
 each evolved using the H{\'e}non orbit-averaged Monte Carlo approach to the evolution of a spherical cluster \citep{Henon1971} using the \texttt{CMC} code \citep{2000ApJ...540..969J, 2003ApJ...593..772F, Rodriguez2022:cmc}.\footnote{There are four simulations with $N=3.2 \times 10^6$ in \texttt{CMC}, which we neglect.} \texttt{CMC} is parallelised for computational efficiency, allowing simulations of interactions involving up to $10^8$ bodies \citep{Pattabiraman2013}.
Stellar and binary evolution in \texttt{CMC} is modelled using \texttt{COSMIC} \citep{COSMIC}, which uses updated binary stellar evolution recipes from \citet{Hurley2002}.

When bodies become close enough that their local gravitational influence on each other decouples them from the overall cluster evolution, that system is evolved through direct integration of small-$N$ resonant encounters using the \texttt{Fewbody} code, which is updated to include the influence of GW emission \citep{FregauRasio2007, Rodriguez2018:formationmassesmergerrates}. Tidal stripping is calculated via a prescription \citep[see][]{Chatterjee2010, Pattabiraman2013} that matches the calculated mass loss rates from GCs based on the N-body simulations of \citet{BaumgardtMakino2003}. 


Hierarchical triples are expected to form via frequent binary-mediated interactions in the cores of GCs \citep[e.g.,][]{SigurdssonHernquist1993, Wen2003}. Although calculation of the secular dynamics of these triples is beyond the scope of \texttt{CMC} (triples are immediately broken apart upon formation during \texttt{CMC} simulations), these triples can be modelled in post-processing. Recent studies have shown these GC triples may play a role in formation of various types of merger events  with rates comparable to that of eccentric captures during
binary-single encounters \citep{Antonini2016,Fragione2020_GCtriples,Martinez2020}.

The \texttt{CMC Catalog} simulations approximately represent the GCs observed in the Milky Way, with a similar range of cluster masses, radii, metallicities, and distance from the centre of the Galaxy after $\sim10$ to $13$~Gyr of their evolution \citep{Kremer2020}. 
The selection of clusters in the \texttt{CMC Catalog} exists on a $4 \times 4 \times 3 \times 3$ grid corresponding to different initial conditions. Clusters are initialised with 2, 4, 8 or 16 $\times 10^5$ particles. The initial virial radius can be 0.5, 1, 2, or 4 pc. The metallicity can be $0.01$, $0.1$, or $1$ $Z_\odot$, 
and the galactocentric distance can be 2, 8, or 20 kpc. 
A \citet{Kroupa} IMF is assumed, with initial masses ranging between $0.08$ and $150$~M$_\odot$, and the initial binary fraction is assumed to be 5 per cent. Each cluster is evolved to $14$~Gyr unless it is destroyed prior to reaching this age. See \citet{Kremer2020} for a detailed description of the cluster simulations.


\subsubsection{Wide field triples}
\label{sec:wide_triple_channel}

Most stellar systems form in weakly-bound, rapidly-dispersing, low-mass clusters or star associations, and evolve in isolation without significant dynamical interactions with neighbouring systems. We refer to the stars within these systems as \textit{field stars}.
Observations suggest that majority of massive BH-progenitor field stars
reside in triple or higher-order multiple systems \cite[e.g.][]{MoeDiStefano2017, Offner2023}. Three-body interactions likely dominate the evolution of a substantial fraction of BH progenitors \citep[see, e.g.,][]{2022MNRAS.516.1406S,Kummer2023} and play a key role for several proposed formation channels of merging BBHs. 

While there are several different ways that three-body interactions can produce merging BBHs \citep[see, e.g.,][]{ Antonini2014, SilsbeeTremaine2017, Antonini2017, RodriguezAntonini2018,LiuLaiu2018,  Fragione2019, FragioneKocsis2020, VignaGomez2021, Stegmann2022, man22, Dorozsmai2024:chetriples, Kummer2024, vg25}, we focus exclusively on triples expected to form GW sources with detectable eccentricities at significant rates. Specifically, we study dynamically-isolated hierarchical triples, in which the inner binary is sufficiently wide that no mass exchange takes place.
This ensures that relativistic precession remains weak, and does not suppress von Zeipel-Lidov-Kozai \citep[ZLK;][]{vonZeipel1924, Lidov1962, Kozai1962} oscillations by the time the inner binary forms a BBH. 
Among these triples, systems in which the the octupole term in the perturbing Hamiltonian \citep[see, e.g.,][]{Ford2000, Ford_2004, Naoz2013} or non-secular perturbations \citep[][]{Antonini2010, KatzDong2012, Antonini2012,
SetoNaoki2013MNRAS.430..558S,
BodeWegg2014, Antognini2014, Antonini2014, Antonini2016, luo16, Grishin17, Grishin2018, Grishin24} become significant over time can reach extremely high eccentricities due to three-body dynamics. These systems can form GW sources that retain detectable eccentricities when they enter the detection band of ground-based GW detectors \citep[][]{Antonini2014, Antonini2017}.
Henceforth, we refer to this formation path as the \textit{wide field triple channel}. We note that in this scenario, we do not consider perturbations by the galactic tides or fly-by stars \citep[for those effects, see e.g.][]{ MichaelyPerets2020, grishin22, steg24}.

Extremely high eccentricities in the inner binary are often associated with non-secular perturbations. We are therefore especially interested in triples that transition to non-secular configurations while remaining dynamically stable. In such non-secular triples, changes in orbital parameters may occur on timescales shorter than the period of the inner and/or outer binary \cite[see e.g.]{KatzDong2012, Antonini2014}.
Given the importance of distinct dynamical regimes for the systems we are interested in, we introduce the following two criteria.

First, we assume that a triple is dynamically stable, if the ratio of the outer and inner semimajor axis is larger than \citep{MardlingAarseth2001}: 
\begin{equation}
\label{eq:dyn_stab}
\left(\frac{a_{\rm out}}{a_{\rm in}}\right)_{\rm crit} = \frac{2.8}{1 - e_{\rm out}}\left(1 - \frac{0.3i}{\pi}\right) \left(\frac{(1 + q_{\rm out}) (1 + e_{\rm out)}}{\sqrt{1 - e_{\rm out}}}\right)^{2/5},
\end{equation}
where $a_{\rm out}$, $a_{\rm in}$ are the outer and inner semimajor axis, respectively; $q_{\rm out} = m_{\rm out}/(m_1 + m_2)$ is the outer mass ratio; and $i$ is the mutual inclination. Dynamically unstable systems typically eject one of the components \citep[see, e.g.,][]{Kiseleva1994,Iben1999,PeretsKratter2012, Stone2019} though the timescale over which this occurs can vary significantly if system parameters are changed \citep[see, e.g.,][]{MushkinKatz2020, Ginat2021, Toonen2022, hayashi22, Zhang2023, Bruenech2024}; therefore, there are different criteria for dynamical stability in the literature \citep[see, e.g.,][]{Eggleton1995, Vynatheya2022, Tory2022}. 

Second, we define the non-secular evolution regime following \citet{Antonini2014}. We consider the evolution of the triple to be non-secular if the angular momentum of the inner binary changes by an order of magnitude within its orbital period due to three-body dynamics \citep[see also ][]{KatzDong2012}: 
\begin{equation}
\label{eq:non_secular_crit}
    \sqrt{1 - e_{\rm in}}\leq \sqrt{1 - e_{\rm crit}} = 5\pi q_{\rm out} \left(\frac{a_{\rm in}}{a_{\rm out} (1 - e_{\rm out})}\right)^3.
\end{equation}

We use the rapid triple population synthesis code, $\texttt{TRES}$ \citep{Toonen2016}, to evolve triples in the secular regime. $\texttt{TRES}$ determines the secular triple orbital evolution while simultaneously incorporating stellar evolution and interaction processes.
For stellar evolution, $\texttt{TRES}$ uses the rapid population synthesis code $\texttt{SeBa}$ \citep[][]{Portegies_Zwart1996, Toonen_2012}, based on fitting \citet{Hurley2000} to the stellar tracks of \citet{Pols1998}. 

Once a triple enters the non-secular regime (see Equation \ref{eq:non_secular_crit}), we follow \citet{Antonini2017} and switch too using the N-body code integrator \texttt{AR-CHAIN} \citep[see][]{MikkolaMerritt2006:algorithmic_regularization_with_velocity-dependent_forces, MikkolaMerritt2008:implementing_few-body_algorithmic_regularization_with_post-newtonian}.
\texttt{AR-CHAIN} solves the equations of motion with post-Newtonian corrections up to 2.5pN, where the  2.5pN term is the dissipative term responsible for energy loss due to GW emission, using algorithmically regularised chain structure \citep[see, e.g.,][]{MikkolaAarseth1989:A_chain_regularization_method_for_few-body_problem, MikkolaAarseth1993:an_implementation_of_N-body_chain_regularization, MikkolaTanikawa:algorithmic_regularization_of_the_few-body_problem} and a time-transformed leap-frog scheme \citep[see, e.g.,][]{Mikkola_Aarseth2002:a_time-transformed_leapfrog_scheme}. 

We simulate $10^6$ triples over grid that is roughly uniform in $\log_{10} Z$ in the range $0.01 Z_\odot \leq Z \leq 1 Z_\odot$. 
We evolve systems from zero-age main sequence (ZAMS) until either: (i) a Hubble time (which we assume to be $13.5\,\rm{Gyr}$) passes; (ii) any of the stars fill their Roche lobe; (iii) the triple transitions to dynamical instability (see Equation~\ref{eq:dyn_stab}); (iv) the triple transitions into the non-secular phase (see Equation \ref{eq:non_secular_crit}); or (v) any of the stars become unbound from the triple system due to a natal kick received by a newly formed compact object.

We exclude systems undergoing mass transfer, because eccentric mass transfer phases are not yet implemented in \texttt{TRES}\footnote{We note that about  67 per cent of all triples is expected to undergo mass exchange see e.g. \citet{Kummer2023}}. However, significant mass exchange is expected to occur with eccentric inner orbits \citep[see, e.g.,][]{Kummer2023}. Reliable modelling of the long-term orbital evolution of triples due to eccentric mass transfer episodes is challenging \citep[but see][]{ Sepinsky2007_Lagrangian_points_eccentric,Sepinsky2007_conservative_eccentric, Sepinsky2009_nonconservative_eccentric, Dosopoulou_2016, Dosopoulou_2016A_II, Hamers_2019_ecc_mt, Rocha2024}. We note that the orbit typically shrinks during mass transfer episodes, increasing the apsidal precession and thereby potentially suppressing three-body dynamics \citep[e.g.,][]{Holman1997,Blaes2002,Liu2015}. This hints that these systems may rarely produce merging BBHs with detectable eccentricities. However, this reasoning may no longer hold if black hole progenitors receive strong natal kicks, which could reestablish a weak hierarchy in a significant fraction of triple systems. Clearly, further work is needed to explore these systems.



Our initial inner binary parameter distributions are motivated by results from recent spectroscopic and long-baseline interferometric surveys of massive binaries found in young, open star clusters and star associations \citep[e.g.][]{KobulnickyFryer2007, Sana2012, Kiminiki2012, DucheneKraus2013ARA&A..51..269D, Sana2014, Kobulnicky2014}.  We draw properties of the outer binary from the same distributions we use for the inner binaries. 
Observations of hierarchical multiple systems of Galactic solar-type stars support this assumption \citep{Tokovinin2014, Tokovinin2006}, as does the recent study by \citet{Shariat2025} motivated by Gaia observations, although it remains uncertain if this holds for massive hierarchical triples.

We assume that the ZAMS masses of primary (most massive of the binary components) stars follow the \citet{Kroupa} power-law distribution in our sampled mass range $20$-$100$\,M$_{\odot}$,  $N$$\sim$$M_{\rm 1, ZAMS}^{-2.3}$. 
The inner and outer mass ratios are uniformly within an interval of $[0.1,1]$ in both cases, broadly consistent with observational studies \citet{KobulnickyFryer2007, Sana2012, Kobulnicky2014}.
For both $a_{\rm in,ZAMS}$ and $a_{\rm out,ZAMS}$, we follow \citet[][]{OpikLaw1924} and sample from log-uniform distributions in an interval [$11.6~\rm{au}$, $4650.5~\rm{au}$].
We discard triple systems that are dynamically unstable at ZAMS according to Equation \ref{eq:dyn_stab}.
We assume that both $e_{\rm in, ZAMS}$ and $e_{\rm out, ZAMS}$ are thermally distributed \citep[see, e.g.,][]{Heggie1975}. The mutual inclination is uniform in cosine. We note that for ultra-wide binaries ($a \gtrsim 1000\ \rm au$) the eccentricity distribution may be super-thermal \citep{hwang22}, and ultra-wide binaries \citep{steg24} and  triples \citep{grishin22} $(a_1\ \rm{and}\ a_2 \gtrsim 5000\ \rm au$, respectively) have a larger parameter space for inducing high inner eccentricity and mergers due to galactic tides. Including the effects of galactic tides is beyond the scope of the current work.



The implementation of stellar winds is described in \citet{Toonen_2012}. For optically thin line-driven winds, we calculate mass loss rates based on the prescriptions of \citet{Vink2001} and \citet{Nieuwenhuijzen1990}. For dust-driven winds, we follow \citet{Reimers1975}, \citet{Nieuwenhuijzen1990} and \citet{Vassiliadis1993}. 
We assume a constant mass loss rate of $1.5\times10^{-4}\,M_{\odot}\rm{yr}^{-1}$ for stars that have crossed the Humphreys-Davidson limit, following \citet{Belczynski2010}. For stripped helium stars, we apply the empirical form of \citet{Hamann1995} with a clumping factor of $\eta = 0.5$ \citep{Hamann98} and a metallicity scaling of $\dot{M}_{WR}\sim Z^{0.86}$ \citep{VinkdeKoter05}. For a more detailed description of the implementation of our stellar wind model, see \citet{Dorozsmai2024:chetriples}.

We compute the BH mass using the ``Delayed'' supernova model of \citet{Fryer_2012}.
 Following \citet{Belczynski2020:evolutionary_roads_leading_to}, we assume that direct-collapse BHs experience spherically symmetric neutrino losses of 1 per cent of the pre-collapse mass, implying that the system only receives a Blaauw kick \citep[e.g.][]{Blaauw1961}. These assumptions broadly agree with the observationally motivated constraints of \citet{Vigna-Gomez2023}. We note that this process has significant uncertainties, and different population synthesis codes apply different assumptions \citep[e.g.,][]{Fryer_2012, Riley2022:compas_paper, Stevenson2019, MandelMuller2020}. 

The natal kick velocity for BHs is calculated as
\begin{equation}
\label{eq:kick}
    v_{\rm BH} = (1 - f_{\rm b})\left(\frac{M_{ \rm NS}}{M_{\rm BH}}\right)v_{\rm kick}, 
\end{equation}
where $f_{\rm b}$ is the fallback fraction \citep{Fryer_2012}, $M_{\rm NS}$ is the canonical neutron star mass ($M_{\rm NS} = 1.4 M_{\odot}$) and $v_{\rm kick}$ is a random kick velocity drawn from the distribution inferred by \citet{Verbunt2017} from proper motion measurements of pulsars. 
The impact of natal kick on the inner and outer orbit due to the core collapse of any of the three stars is determined according to \citet{Pijloo2012}.

Orbital evolution due to three-body dynamics is determined via the secular evolution equations with terms including the quadrupole \citep{Harrington1968} and octupole terms (\citealt{Ford_2004} with corrections of \citealt{Naoz2013}) from Hamiltonian perturbation theory \citep[e.g.][]{Valtonen2006}. 
Additional non-Keplerian forces are also taken into account, such as the additional apsidal precession from tidal and rotational bulges and general relativity \citep{SmeyersWillems2001, Blaes2002, Liu2015} and tidal dissipation \citep{Hurley2000} and GW emission \citep{Peters64}.
The impact of stellar winds on the inner and outer orbit is determined by assuming fast adiabatic winds \citep[e.g.,][]{Toonen2016}. 
This results in a set of first-order ordinary differential equations for the system parameters \citep[see Eq. 39 in][]{Toonen2016}.

\subsection{Differences in models of stellar evolution in \texttt{CMC} and \texttt{TRES}}

Some aspects of the models chosen for the core collapse of massive stars differ between \texttt{CMC} and \texttt{TRES}. The assumed core-collapse mechanism plays a critical role in determining remnant mass and can significantly alter the orbit of the stellar systems and the structural evolution of the GC due to the associated instantaneous mass loss and natal kicks. We outline these model differences below and summarise their expected impact on our results.

In contrast to the \textit{Delayed} supernova prescription applied in our triple simulations, \texttt{CMC} adopts the \textit{Rapid} model of \citet{Fryer_2012}. 
This affects the evolution of systems containing stars with initial masses in the range of $M_{\rm ZAMS}\approx$ 20-35$\,M_{\odot}$ at solar metallicity. The affected mass range becomes smaller with decreasing metallicity, e.g. $M_{\rm ZAMS}\approx20$--$22$\,M$_{\odot}$ at $Z = 0.006$ \citep[see  Fig. 11 in ][]{Fryer_2012}. Outside this mass range, the differences are negligible. A distinctive consequence of the \textit{Rapid} prescription is that `lower mass gap' ($2$--$5\ M_\odot$) compact objects cannot form 
\citep{Zevin2020}. This also implies reduced mass ejection for stars undergoing core-collapse in the affected mass range, compared to the \textit{Delayed} model. If \texttt{TRES} utilised the \textit{Rapid} prescription as well, we would expect fewer disrupted triples, more inner binaries with shorter orbital periods, and potentially higher formation efficiency of GW sources within the affected mass range. Thus, the \textit{Delayed} model is the more conservative assumption for triples. 

In the \texttt{CMC} simulations, the kick velocity is calculated as 
\begin{equation}
\label{eq:cmc_kick}
    v_{\rm BH} = (1 - f_b)v_{\rm NS},
\end{equation}
where $v_{\rm NS}$ is the kick velocity of a neutron star drawn from a Maxwellian distribution with dispersion $\sigma = 265$ km s$^{-1}$ \citep{Hobbs2005}. 
\texttt{TRES} uses the mixture model of \citet{Verbunt2017} , which comprises two Maxwellians, $\sigma_1 = 75$ km s$^{-1}$ and $\sigma_2 = 316$ km s$^{-1}$ and furthermore, the kick velocity is also scaled with the BH mass, see equation \ref{eq:kick}. We explore the impact of the different SN and natal kick models adopted in \texttt{CMC} and \texttt{TRES} in appendix \ref{appendix:impact_sn_kick}.


\subsection{Population of merging binaries at different redshifts}
\label{subsec:weighing_scheme}
In order to determine the demographics of merging BBHs across different redshifts, we apply the following weighting scheme to the simulation data. 
We transform the synthetic populations introduced in the preceding section into distributions that reflect the statistics of BBH mergers occurring at a given redshift, $z_{\rm obs}$ by assigning each GW source $i$ a weight $w_i$ proportional to its predicted merger rate. This weight, in turn, is proportional to the metallicity-dependent star formation rate density, $\rm{SFRd}$, in the specific environment (i.e., GC or field) at the redshift at which the progenitor stars were formed, $z_{\rm form}$, and at the metallicity of the progenitor system, $Z$. The assigned weight for the $i$th GW source is 
\begin{equation}
\label{eq:weighing_scheme_general}
    w_i = k_i \frac{ {\rm SFRd}(z_{{\rm form}, i}, Z_i)}{\hat{M}}.
\end{equation}
Since $\rm{SFRd}$ is expressed in terms of stellar mass, we divide it by $\hat{M}$, the average ZAMS mass of stellar systems (single, binary or triple) to determine the average number of stellar systems formed per unit time and comoving volume. The term $k_i$ represents the occurrence rate of the specific GW source, expressed as the fraction of all stellar systems formed at $z_{\rm form}$. Thus, through Equation \ref{eq:weighing_scheme_general}, we obtain the merger rate density corresponding to the given merging BBH. The sum of the weights $w_i$ gives the total merger rate density for a particular formation channel at the observed redshift, $z_{\rm obs}$. The distributions obtained with this weighting scheme are thus normalised to the total merger rate density of the given formation channel.

Since the choice of $k_i$ depends on the channel, we discuss the remaining details of the weighting scheme for GCs and triples separately.



\subsubsection{Globular clusters}
The \texttt{CMC Catalog} is defined on a grid that is uniformly spaced in $\rm{log}(M_{\rm cl})$ and $\rm{log}(r_{\rm vir})$. We assign weights to each \texttt{CMC} GW source depending on the properties of the host GC to ensure that the predicted GW population reflects our assumptions regarding the initial GC properties, which are detailed below. We assume an initial cluster mass function (CIMF) that follows a Schetcher-like function,
\begin{equation}
\label{eq:CIMF}
    N_{\rm{CIMF}}(M_{\rm cl}) \propto \left(\frac{M_{\rm cl}}{M_*}\right)^{-2} {\rm{exp}}\left(\frac{-M_{\rm cl}}{M_*} \right),
\end{equation}
with $M_{\rm cl}$ in the range $10^4$-$10^8 \, M_{\odot}$.
We adopt a cut-off mass of $M_* = 10^{6.3}\, M_{\odot}$ \citep[see, e.g.,][]{AntoniniGieles2020, FishbachFragione:2023:GCs}. The functional form of Equation \ref{eq:CIMF} is motivated by observations of young stellar clusters \citep[see, e.g.,][]{ZhangFall1999, LadaLada2003, Larsen2009, Portegies_Zwart2010}. We assume $r_{\rm vir}$ follows a log-uniform distribution on a interval of $[0.5, 4]$ pc \citep[but see, e.g.,][ for different assumptions]{ FishbachFragione:2023:GCs}. We assume that initial cluster masses and virial radii are uncorrelated and their distributions do not depend on redshift or metallicity.  

The initial masses of the clusters in the \texttt{CMC Catalog} range from $1.2\times10^4$ to $9.6\times10^5\,M_{\odot}$. However, observations of Milky Way GCs indicate that clusters exist outside of this mass range. Additionally, the number of BBH mergers is predicted to scale with the initial cluster mass as $M_{\rm cl}^{1.6}$ over a cluster mass range of $10^2$--$10^6\,M_{\odot}$, for a given initial virial radius \citep[][]{AntoniniGieles:2020:cBHBd}. To account for mergers from clusters not covered by the \texttt{CMC Catalog}, we apply a correction method similar to that of \citet{Kremer2020} and \citet{FishbachFragione:2023:GCs}. Specifically, we multiply the computed merger rate by a correction factor, $f_{\rm cor}$, that estimates the contribution of GW sources from the missing clusters. We assume that the number of mergers from a cluster with $M_{\rm cl}$ and $r_{\rm vir}$ can be approximated with $N_{\rm GW} = K(r_{\rm vir})M_{\rm cl}^{1.6}$, where $K(r_{\rm vir}) = \rm{exp}(0.056 r_{\rm vir}^2 - 0.7111 r_{\rm vir} -15.643)$ as derived from the \texttt{CMC Catalog} and $r_{\rm vir}$ is in the units of $\rm{pc}$, while the units of $M_{\rm cl}$ are in $M_{\odot}$. The fitting formula is valid for $0.5\leq r_{\rm vir}/\rm{pc} \leq 4$. We note that this extrapolation method accounts only for the change in the merger rate and neglects how other properties of merging binary black holes, such as their eccentricity distributions, may vary with increasing $M_{\rm cl}$. Therefore, we assume that the eccentricity distribution remains unchanged with the inclusion of more massive clusters not represented in the \texttt{CMC Catalog}. However, this assumption is not strictly valid. In particular, the shape of the eccentricity distribution depends on the initial number density of the cluster, as shown in Appendix \ref{appendix:changing-GC-e-with-rv}. If $r_{\rm vir}$ and $M_{\rm cl}$ are uncorrelated, then more massive clusters tend to have higher number densities on average, which could lead to different eccentricity distributions. However, we do not expect this would change our conclusions significantly, as the fraction of GW sources with eccentricities at 10 Hz detectable by XG detectors ($e_{\rm 10Hz}\approx 10^{-3}$-$10^{-4}$, see discussion later) does not vary substantially with increasing number density (see Fig. \ref{fig:density_profile_rvcomp}).


Integrating $N_{\rm GW}$ over the CIMF and the initial $r_{\rm vir}$ distribution function gives the total number of GW sources from a population of GCs, that is 
\begin{equation}
    N_{\rm GW, tot} = \int_{M_{\rm cl, min}}^{M_{\rm cl, max}}\int_{r_{\rm min}}^{r_{\rm max}} N_{\rm GW} N_{\rm CIMF}N_{\rm r,vir} dM_{\rm CL} dr_{\rm vir},
\end{equation}
where $N_{\rm rvir}$ is the normalised distribution of $r_{\rm vir}$. The correction factor ($f_{\rm cor})$ is then calculated by dividing the total number of mergers computed with the assumed complete initial cluster mass range, $M_{\rm cl, min} = 10^4$, $M_{\rm cl, max}$ = $10^8 \, M_{\odot}$, with the total number of mergers calculated with the mass range of the \texttt{CMC Cluster Catalog}.

Finally, we compute the merger rate density of GW sources from the GC channel as:
\begin{equation}
    \label{eq:merger_rate_density_GC}
     R_{\rm GC}(z_\mathrm{obs}) = \frac{1}{\hat{N}_{\rm tot}}f_{\rm cor}\sum\limits_i^{N_{\rm GW}} \frac{{\rm SFRd}(z_{\mathrm{form},i} , Z_i)}{\hat{M}} w_{{\rm cl},i},
\end{equation}
where $z_\mathrm{obs}$ is the redshift at which the observed merger occurs, $N_{\rm GW}$ is the total number of GW sources in the dataset and $w_{\rm cl}$ is the weight representing the occurrence rate of the host cluster of the specific merging BBH. We compute $w_{\rm cl}$ by integrating the PDFs corresponding to $M_{\rm cl}$ and $r_{\rm vir}$ and taking their product. For integration boundaries, we choose the midpoints of the neighbouring grid values in logarithmic space. $\hat{N}_{\rm tot}$ represents the total initial number of stars in the \texttt{CMC Cluster Catalog}, corrected by the initial cluster weights, i.e. $\hat{N}_{\rm tot} = \sum_i^{N_{\rm tot}} w_{{\rm cl},i}$, where $N_{\rm tot}$ is the total initial number of stars in the catalogue. From Equation \ref{eq:merger_rate_density_GC}, it follows that $k_i = f_{\rm cor} w_{\rm cl}/\hat{N}_{\rm tot}$ for the GC channel. 

To determine the star formation rate in GCs, we use the phenomenological fit of \citet{RodriguezLoeb2018} based on the GC formation model of \citet{ElBadry2019}, which expresses the stellar mass formed in GCs at a given redshift in units of $M_{\odot}\rm{yr}^{-1}Mpc^{-3}$. 

We renormalise the GC formation model of \citet{ElBadry2019} such that integrating it over cosmic time yields the initial GC mass density ($\rho_{\rm GC,i}$). Here $\rho_{\rm GC,i}$ is essentially the mass density of GCs that would be present today in the absence of GC evaporation \citep[see e.g. eq. 8 in][]{AntoniniGieles2020}.
We relate $\rho_{\rm GC,i}$ to the present-day mass density by a constant factor representing GC evaporation, $K_{\rm ev} = \rho_{\rm GC,i}/\rho_{\rm GC,0}$. We determine $K_{\rm ev}$ following \citet{AntoniniGieles2020}:
\begin{equation}
\label{eq:Kev}
K_{\rm ev} = \frac{\int_{M_{\rm GC,min}}^{M_{\rm GC,max}} N_{\rm CIMF} M_{\rm CL} dM_{\rm CL}}{\int_{M_{\rm GC,0,min}}^{M_{\rm GC,0,max}} N_{\rm CMF, 0} M_{\rm CL,0} dM_{\rm CL,0}},
\end{equation}
where $N_{\rm CMF, 0}$ is the present-day GC mass function, for which we adopt the evolved Schechter function presented in \citet{AntoniniGieles2020}.
The integration limits represent the minimum and maximum GC masses at their initial state and at the present day. We assume $M_{\rm GC,min} = 10^5\,M_{\odot}$, $M_{\rm GC,max} = 10^7\,M_{\odot}$ \citep[see e.g.][for similar assumptions]{RodriguezLoeb2018}, while for the present-day GCs, we adopt $M_{\rm GC,0,min} = 10^{2.3}\,M_{\odot}$ and $M_{\rm GC,0,max} = 10^7\,M_{\odot}$. The value of $M_{\rm GC,0,min}$ corresponds to approximate mass of Koposov 2 \citep[][]{Koposov2007}, which is the least massive GC in the catalog of \citet{Harris2010}.
We then normalize the GC formation model to $\rho_{\rm GC,i} = M_{\rm avg, GC} n_{\rm GC,0}  K_{\rm ev}$ where $M_{\rm avg, GC}$ is the present-day average GC mass, assumed to be $1.5 \times 10^5\,M_{\odot}$ \citep[see, e.g.,][]{AntoniniGieles2020} and $n_{\rm GC,0}$ is the present-day GC number density, for which we adopt a value of  $1.5\,\rm{Mpc}^{-3}$, following \citet{Harris2015}. For our adopted CIMF, we obtain $\rho_{\rm GC,i} = 8.4\times 10^{15}\,M_{\odot}\rm{Gpc^{-3}}$.

Both $n_{\rm GC,0}$ and $K_{\rm ev}$ are uncertain, and different studies have adopted varying values \citep[see, e.g.][]{ElBadry2019, Choksi2019, RodriguezLoeb2018, AntoniniGieles2020}. Additionally, our estimate of $\rho_{\rm GC,i}$ is approximately a factor of three lower than that of \citet{AntoniniGieles2020}. This difference is primarily due to choice of the minimum initial GC mass, which was set to $M_{\rm GC,min} = 10^2\,M_{\odot}$ in \citet{AntoniniGieles2020}. Although, adopting such a low $M_{\rm GC,min}$ increases $\rho_{\rm GC,i}$  by a factor of three and the merger rate scales linearly with $\rho_{\rm GC,i}$ (see equation \ref{eq:merger_rate_density_GC}), this does not lead to a proportional increase in the predicted BBH merger rate. Clusters with low initial masses of $\sim10^2$-$10^3\,M_{\odot}$ are not expected to produce GW sources efficiently. Adopting $M_{\rm GC,min} = 10^2\,M_{\odot}$ assumes that a substantial fraction of $\rho_{\rm GC,i}$ is contributed by such low mass clusters, reducing the mass density available for higher mass clusters which are the primary producers of BBH mergers.

We compute the metallicity-specific star formation rate for GCs by convolving the renormalised star formation history model of \citet{ElBadry2019} with a metallicity distribution of stellar mass formation ($f_{\rm Z})$. We derive $f_{\rm Z}$ by combining the mass-metallicity relation of \citet{LangerNorman2006} and the galaxy stellar mass function (GSMF) of \citet{Pantter2004M}. More specifically, 
\begin{equation}
\label{eq:SFRd}
    {\rm SFR}_{\rm GC}(z,Z) = {\rm SFR}_{\rm GC}^{*}(z)\int\limits_{Z - \Delta Z}^{Z + \Delta Z}  f_{\rm Z^\prime}(z,Z^\prime) {\rm d}Z^\prime,
\end{equation}
where the integration boundaries are set as the midpoints of the neighbouring grid values of $Z$ in the \texttt{CMC Cluster Catalog} in logarithmic space. The $\rm{SFR_{\rm GC}}^{*}$ is the cosmic star formation history in GCs. 
The term $f_{\rm Z}$ is determined as
\begin{equation}
\label{eq:met_dist_synt}
    f_{\rm Z} = k M_s(Z) \frac{dN}{dZ}(Z) = k M_s(Z) \frac{dN}{dM_s} \frac{dM_s}{dZ},
\end{equation}
where $k$ is a normalization constant, $dN/dZ$ is the number of galaxies per metallicity bin and $M_s(Z)$ stellar mass associated with the specific metallicity bin. The term $dN/dM_s$ represents the number density of galaxies per stellar mass bin, as given by the GSMF from \citet{Pantter2004M} and $dM_s/dZ$ is determined from the mass-metallicity relation of \citet{LangerNorman2006}.
For a similar method of determining the metallicity-specific star formation rate, see also e.g. \citet{Neijssel2019} and for an alternative method see \citet{Chruslinska2019}. For simplicity, we assume a one-to-one relation between stellar mass and metallicity. This in in contrast with the common assumption that the metallicity follows a Gaussian distribution for a given galaxy stellar mass \citep[see, e.g.,][where a 0.1 dex spread is assumed]{Tremonti2004}.



\subsubsection{Wide field triples}

Since triple systems are directly sampled from our assumed initial distributions for field stars, we do not need to employ such a complex weighting scheme as required for the GC channel. However, we must account for the fact that our sampling covers only a limited
fraction $f_{\rm pm}$ of the full parameter space of field stars. Bearing this in mind, we compute the merger rate density for this channel as
\begin{equation}
    \label{eq:merger_rate_density_triple}
     R_{\rm triple}(z_\mathrm{obs}) = \frac{f_{\rm pm}}{N_{\rm tot}}\sum\limits_i^{N_{\rm GW}} \frac{{\rm SFR}_{\rm field}(z_{{\rm form},i}, Z_i)}{\hat{M}} ,
\end{equation} 
where ${\rm SFR}_{\rm field}(z_{{\rm form},i}, Z_i)$ 
is the metallicity-dependent star formation rate for field stars, and $\hat{M}$ is the average ZAMS mass of stellar systems. This is determined in essentially the same way as for the GC channel but adopting the cosmic star formation model of \citet{Madau_2017} instead of that of \citet{ElBadry2019}. $N_{\rm tot}$ is the total number of sampled systems.

We determine $\hat{M}$ by adopting the following multiplicity fractions for single, binary and triple stars: $f_{\rm single} = 0.06$, $f_{\rm binary} = 0.21$  and $f_{\rm triple} = 0.73$. These multiplicity fractions are broadly consistent with the observations of \citet{MoeDiStefano2017} and \citet{Offner2023} of O/B stars with $M_{\rm ZAMS}\gtrsim20\,M_{\odot}$, if one assumes quadrupole stellar systems can be modelled as triples \citep[which  might not always be a valid assumption; see, e.g.,][]{Hamers2021:quad, Vynatheya2022:quadruples}.

We compute $f_{\rm pm}$ using a Monte Carlo method based on the following assumptions about the full parameter space of massive triple stars. At ZAMS, stellar triples are assumed to be detached (we ignore contact binaries) and dynamically stable, with inner and outer semimajor axes ranging from $0.005$~au to $10^5$~au. While the same range is assumed for the inner and the outer orbit, the dynamical stability criterion (Equation \ref{eq:dyn_stab}) typically requires the outer orbit to be at least a factor of 3–4 larger than the inner orbit for a given system. We assume $M_{1,\rm{ZAMS}}$ ranges from 0.08 $M_{\odot}$ to 100 $M_{\odot}$, $q_\mathrm{in}$ and $q_\mathrm{out}$ range from 0.1 to 1.0,
 and initial $e_\mathrm{in}$ and $e_\mathrm{out}$ range from 0 to 0.9. We find $f_{\rm pm} \approx 1.1 \times 10^{-4}$.
Our assumption that the tertiary cannot be more massive than the total mass of the inner binary is common in the literature \citep{Toonen2022, Kummer2023, Hamers2022}. Recently, observations of triples with $q_\mathrm{out}>1$ have been made \citep[see e.g.][]{Eisner2022}. Furthermore \citet{Shariat2025} found that a non-negligible fraction of triples have a initially a tertiary companion which is more massive than the inner binary, based on comparisons between triple stellar models and triple systems detected Gaia. We assume that stellar multiplicity fractions do not correlate with other stellar parameters 
\citep[which may not be consistent with observations, e.g.,][]{MoeDiStefano2017}. We compute $f_{\rm pm}$ by first sampling $\sim 10^7$ triple systems from the full parameter space of all triple stars. as defined above, then determining the fraction of systems that fall within the sampling range used for the simulation of the evolution of triple stars.

\begin{table*}

\caption{A summary of the most important features of model variations exploring how sensitive our results are to uncertainties in the initial properties of GCs and the metallicity-specific star formation rate density.  See section \ref{subsec:model_variations} for more details.}
\centering
\resizebox{0.7\textwidth}{!}{%
\begin{tabular}{lccl}
\hline
\textbf{Model name} & \multicolumn{3}{c}{\textbf{Features}} \\ \hline
                      & \textbf{cut-off mass in ICMF} & \textbf{Mass-metallicty relation} & \textbf{Galaxy stellar mass function} \\ \hline
\textit{Fiducial}     & $M_{*}$ = $10^{6.3}\,M_{\odot}$   & \citet{LangerNorman2006}   & \citet{Pantter2004M}          \\
\textit{GCmax}     & $M_{*}$ = $10^{7.5}\,M_{\odot}$   & \citet{Ma2016}             & \citet{Pantter2004M}          \\
\textit{TRIPLEmax} & $M_{*}$ = $10^{5.5}\,M_{\odot}$   & \citet{LangerNorman2006}   &  \citet{Furlong2015}     \\   \hline  
\end{tabular}%
}\label{tab:model_variations}
\end{table*}

\subsection{Model variations}
\label{subsec:model_variations}
There are several uncertainties in the modelling of both channels that affect the properties and rates of merging BBHs. These include: (i) uncertainties related to different aspects of stellar evolution, such as stellar winds (compare e.g., \citealt{Vink2001}, \citealt{Bjorklund2021:new_prescipII};  \citealt{KrtickaKubat2017}, \citealt{Fullerton2006:massloss}), stellar evolution beyond the Humphreys-Davidson limit (see, e.g., \citealt{Humphreys1979}; \citealt{Smith2004}; \citealt{VinkSabhahit2023}; \citealt{Gilkis2021}), supernovae mechanisms (compare e.g. \citealt{Fryer_2012},\citealt{MandelMuller2020}), supernova kicks (compare e.g. \citealt{Hobbs2005}; \citealt{Arzoumanian2002}; \citealt{Igoshev2021}); (ii) uncertainties in star formation rate models and metallicity distributions (compare e.g., \citealt{MadauDickinson2014}, \citealt{Madau_2017}, \citealt{Strolger2004}, \citealt{vanSon2023}, \citealt{Chruslinska2019}) (iii) uncertainties in the formation, initial properties and the evolution of GCs \citep[see, e.g.,][]{Gnedin2014, Forbes2018, Choksi2019}; and (iv) uncertainties in the initial conditions of field stars (i.e. stellar multiplicity fraction, initial properties of triple stars; see \citealt{MoeDiStefano2017}; \citealt{Offner2023}). 

In order to explore how sensitive our results are to uncertainties, we construct three additional model variations, summarised in Table \ref{tab:model_variations}. We choose initial conditions in the first two of our model variations such that the relative contribution of eccentric mergers from the GC channel (\textit{GCmax}) and the wide triple channel (\textit{TRIPLEmax}) are maximised.

In \textit{GCmax}, we assume a CIMF with cut-off mass at $M_* = 10^{7.5}\,M_{\odot}$, and adopt the mass-metallicity relation of \citet{Ma2016}, which on average predicts higher-metallicty star formation at a given redshift than the model of \citet{LangerNorman2006}. Since we assume a different CIMF than in our fiducial model, the initial mass density $\rho_{\rm GC,i}$ changes to  $3.3\times 10^{15}\,M_{\odot}\rm{Gpc}^{-3}$ (see equation \ref{eq:Kev}).
In \textit{TRIPLEmax}, we assume an exponential cut-off in the CIMF at $M_{*} = 2\times10^{5.5}\,M_{\odot}$ following \citet{Gieles2006} and \citet{Larsen2009}, and assume a galaxy stellar mass function (GSMF) of \citet{Furlong2015}, which is based on cosmological hydrodynamical simulations calibrated to the observed present-day galaxy stellar masses and sizes. In contrast to the GSMF of \citet{Pantter2004M}, the model of \citet{Furlong2015} predicts that the GSMF shifts to increasingly lower mass ranges with increasing redshifts, leading to a metallicity-specific star formation model that favours lower metallicities at higher redshifts compared to our fiducial model. In this model variation $\rho_{\rm GC,i} = 1.9\times 10^{16}\,M_{\odot}\rm{Gpc}^{-3}$. 

\section{Evolution of eccentricity distributions and rates with redshift}
\label{sec:ztrend}

\begin{figure*}
\includegraphics[width=\textwidth]{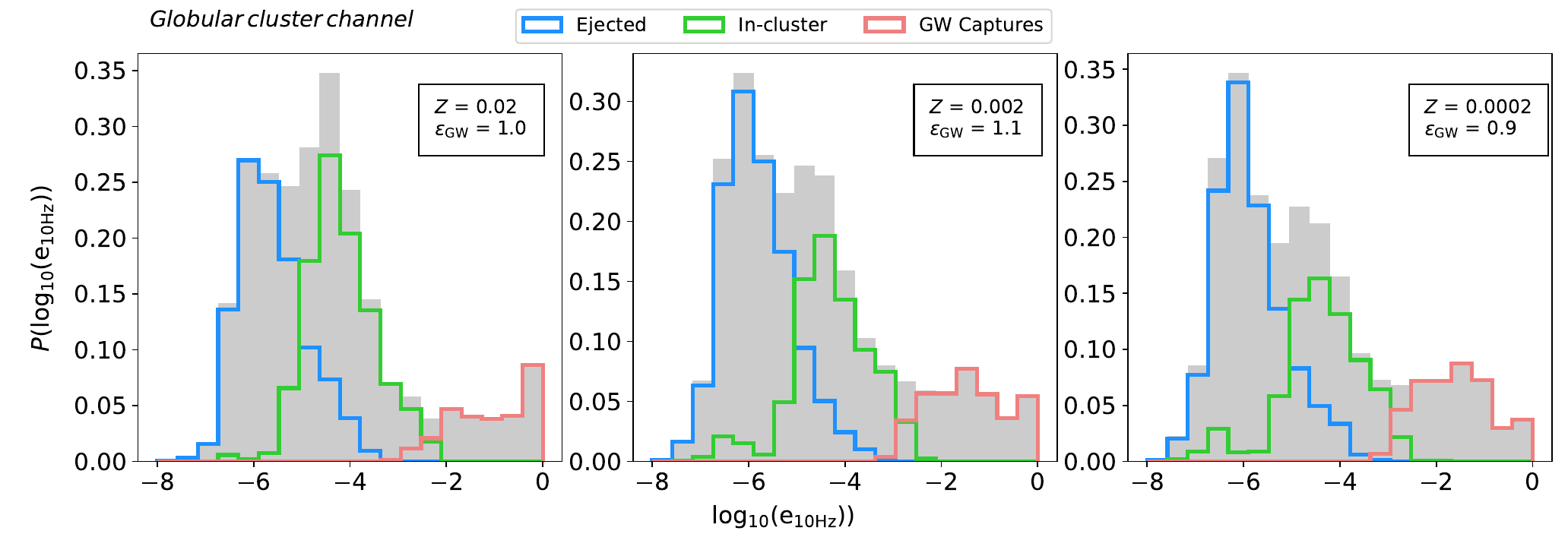}\hfill
\includegraphics[width=\textwidth]{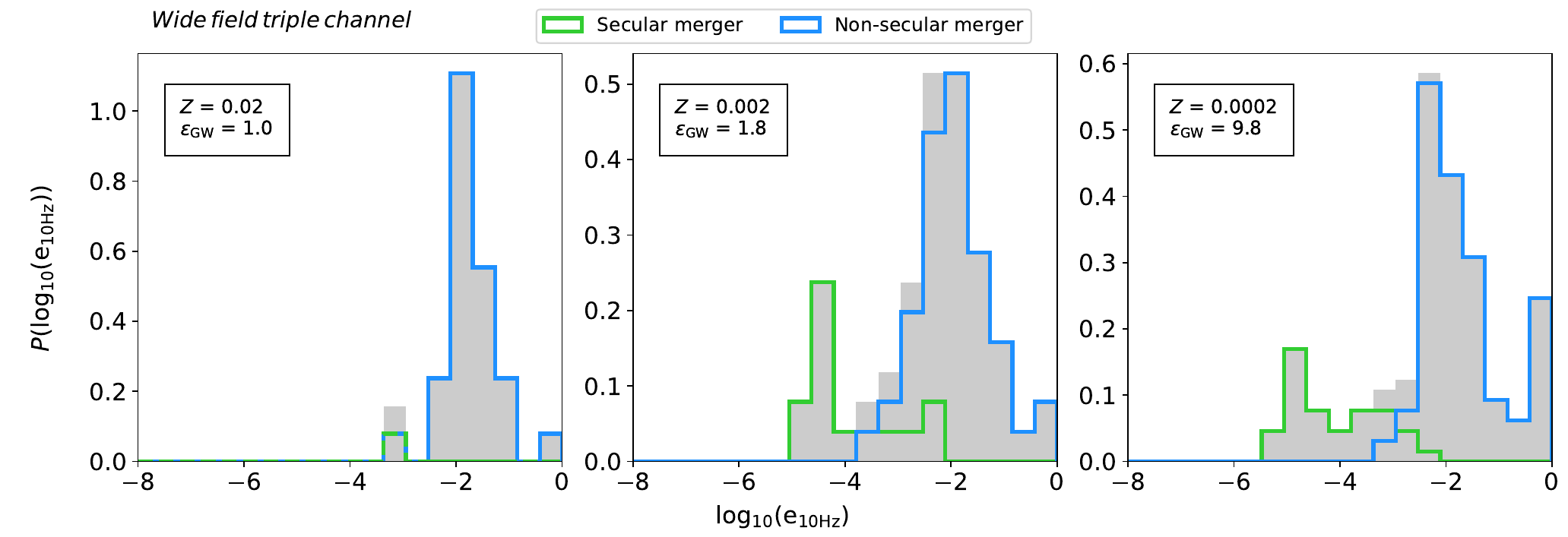}
  \caption{The eccentricity distributions from GCs (top panel) and wide field triples (bottom panel) at three different metallicities, $Z$. In both cases we highlight the contributions from different sub-channels. For the GC channel, we distinguish the processes introduced in detail in Section \ref{sec:GC_channel}. For field triples, we distinguish sources that have experienced non-secular evolution, and systems in which purely secular processes led to the merger of the inner BBH. In all panels the grey histograms, showing the entire contribution from one channel, have been normalised to unity. The formation efficiency of GW sources relative to that at $Z=0.02$, $\epsilon_{\mathrm{GW}}$, is shown in each panel.
  }
  \label{fig:GCFT_Z}
\end{figure*}

\subsection{Influence of metallicity
}
\label{sec:Ztrend}

The evolution of massive BBH progenitor stars is highly sensitive to metallicity. While metallicity evolves with redshift, other progenitor properties - e.g., stellar mass distribution, host cluster characteristics, and initial triple conditions - are fixed in our models. Consequently, the cosmic evolution of massive star metallicity is a primary driver of the redshift dependence in merging BBH demographics.

In Fig. \ref{fig:GCFT_Z}, we show the eccentricity distribution of GW sources from the GC channel (upper panel) and the wide field triple channel (lower panel) when the systems are initialised at three different metallicities: $Z = 0.02$, $0.002$ and $0.0002$. We indicate $\epsilon_{\mathrm GW}$, which is the BBH formation efficiency of each channel at the given metallicity relative to the formation efficiency of the same channel at solar metallicity in the relevant panel (we assume $Z_{\odot} = 0.02$). We define formation efficiency as the number of merging BBHs per simulated stars. Therefore $\epsilon_{\mathrm GW} = f_\mathrm{mergers}(Z) / f_\mathrm{mergers}(Z = Z_\odot)$, where $f_\mathrm{mergers}(Z)$ is number of mergers as a fraction of total number of stars simulated at $Z$. The  parameter $\epsilon_{\mathrm GW}$ therefore indicates how sensitive the merger rate density is to changing metallicity. 


For the GC channel, the formation efficiency of BBHs changes only moderately with metallicity, in agreement with earlier studies \citep[e.g.,][]{AntoniniGieles2020, Mapelli2022}. The shape of the eccentricity distribution is also a weak function of metallicity. These trends are perhaps somewhat unexpected, as the mass-loss rates of massive stars due to line-driven winds depend strongly on metallicity \citep{Vink2001}. In principle, this affects several aspects of the evolution of massive stars and their host GCs, including the structural evolution of GCs, and the mass spectrum and natal kick velocities of BHs. The latter also determines initial number density of BHs in the dense BH-dominated core of GCs, $n_\mathrm{BH}$ \citep[see, e.g., Fig. 3 in][]{Kremer2020}. As we discuss in more detail in Appendix \ref{appendix:changing-GC-e-with-rv}, $n_\mathrm{BH}$ determines the rate of strong (resonant) encounters, and, therefore, the formation efficiency of GW sources within GCs. Additionally, the formation efficiency of each sub-channel (ejected, in-cluster, or GW capture) changes at a different rate with changing $n_\mathrm{BH}$. Thus changing $n_\mathrm{BH}$ changes formation efficiency and the shape of the eccentricity distribution of merging BBHs (see e.g. Fig. \ref{fig:e_GC_rv}). However, as it turns out, metallicity has a remarkably minor effect on the $n_\mathrm{BH}$, and therefore does not significantly affect the dynamics of the BH subsystem (compare e.g. Fig. \ref{fig:density_profile_gc} with Fig. \ref{fig:density_profile_rvcomp}).


In fact, the relative minor impact of metallicity is due to the less massive BHs that form at higher metallicities, rather than due to the moderate change in $n_{\rm BH}$. As shown in Fig. \ref{fig:GCFT_Z}, the relative rate of in-cluster BBH mergers increases with metallicity, while the rate of ejected mergers decreases. When BBHs undergo dynamical encounters, they are imparted a recoil velocity that can eject them from the cluster if it exceeds the cluster escape velocity $v_\mathrm{esc}$. The recoil velocity magnitude is of the order of the orbital velocity, $v_\mathrm{orb} \approx \sqrt{GM_{\rm bin}/a}$, where $M_{\rm bin}$ is the BBH total mass. For smaller masses, smaller separations $a$ are required for $v_\mathrm{orb} \gtrsim v_\mathrm{esc}$; that is, lower-mass binaries (more commonly found at higher $Z$) require more hardening than higher-mass binaries in order to be ejected from the cluster. The GW inspiral timescale is \citep{Peters64, Kremer:2021:WDs}:

\begin{equation}
    t_\mathrm{insp} \approx 0.1~\mathrm{Myr} \left(\frac{a}{0.02 \mathrm{au}}\right)^4\left(\frac{m}{10~M_\odot}\right)^{-3}\left(1-e^2\right)^{\frac{7}{2}},
\end{equation}
while the ejection timescale is \citep[see also][]{Samsing17}:
\begin{equation}
    t_\mathrm{ej} \approx 9~\mathrm{Myr} \left(\frac{\sigma_v}{10~\mathrm{km ~s}^{-1}}\right)\left(\frac{v_\mathrm{esc}}{40~\mathrm{km ~s}^{-1}}\right)\left(\frac{n_\mathrm{BH}}{10^6~\mathrm{pc}^{-3}}\right)^{-1}\left(\frac{m}{10 M_\odot}\right)^{-2},
\end{equation}
where $m = m_1 = m_2$ represents one component of an equal-mass binary.
Because $t_\mathrm{insp} \propto a^4 m^{-3}$ while $t_\mathrm{ej} \propto m^{-2}$, lower-mass BBHs often have shorter inspiral timescales than ejection timescales. For a smaller $M_{\rm BH}$, $a$ has to be smaller for the BBH to get ejected; therefore, it hardens for longer inside the cluster, increasing its likelihood of undergoing interactions that would lead to a merger due to GWs. 

We confirm that these trends are also seen in simulations produced with the fast, BBH-focussed, efficient and approximate cluster simulation code \texttt{cBHBd} \citep{AntoniniGieles:2020:cBHBd} over a logarithmically-spaced grid of 25 metallicities that span between and beyond the full range represented by \texttt{CMC}. In both simulations, the relative contributions of in-cluster and ejected mergers increase and decrease, respectively, with increasing metallicity.



The eccentricity distribution and formation efficiency of GW sources from wide field triples are more sensitive to metallicity than those of the GC channel. Notably, the peak in the highest-eccentricity bin becomes more prominent at lower $Z$. Additionally, the formation efficiency  increases as $Z$ decreases. Increased mass loss rates of line-driven stellar winds at higher metallicities lead to generally wider orbits 
and to lower CO core masses \citep[e.g.,][]{Belczynski2010}. Triples with such characteristics become unbound more efficiently due to BH natal kicks, leading to fewer bound BH triples, and consequently to a lower formation rate of GW sources at higher metallicities.

At solar metallicity, only $2.5$ per cent of our systems form BH triples, while this percentage is $19.4$ per cent at $Z = 0.01~Z_\odot$; in both cases, about $80$ per cent of are dynamically stable. 
We find that most GW sources originate from BH triples with weak hierarchy, $a_{\rm in}/(a_{\rm out}(1-e_{\rm out})) \gtrsim 0.1$ \citep[see also][]{Antonini2014}. 
Triples with weak hierarchy and relatively short inner orbits ($a_{\rm in} \lesssim 10^{4.25}\,R_{\odot}$) are efficiently formed at low metallicity but are absent at $Z=Z_\odot$ due to stronger stellar winds, 
further explaining the decrease in formation efficiency at higher metallicity. 

Qualitatively similar but quantitatively different trends were found by \citet{RodriguezAntonini2018} and \citet{Antonini2018} regarding the formation efficiency of GW sources from triples. Those studies find an increase of a factor of 100 in formation efficiency over the same range of metallicity as shown in Fig \ref{fig:GCFT_Z}. 
This difference are likely due to different modelling choices: for example, \citet{RodriguezAntonini2018} samples triples from a more extended parameter space and does not exclude systems in which Roche-lobe overflow occurs in the inner binary. However, they neglect three-body dynamics until the inner binary forms BHs, which might not be a valid assumptions for a significant fraction of systems \citep[see][]{Kummer2023}.



Since BBH formation efficiency in GCs is weakly dependent on metallicity, the merger rate density from these environments is expected to steadily follow the cosmic SFR in GCs, albeit with a delay time. 
This supports ideas that XG ground-based GW detectors could play a unique role in constraining the cosmic GC formation rate and revealing key information about GC formation and evolution \citep{Romero-Shaw:2021:GCs, FishbachFragione:2023:GCs}.

We note that BBH formation efficiency estimates are subject to several uncertainties \citep[see, e.g.,][]{AntoniniGieles2020}, and present the influence of these on the merger rates from both channels in Section 
\ref{sec:uncertainties}.


\subsection{Detected eccentricity}
\label{sec:detected_eccentricity}

\begin{figure*}
\includegraphics[width=\textwidth]{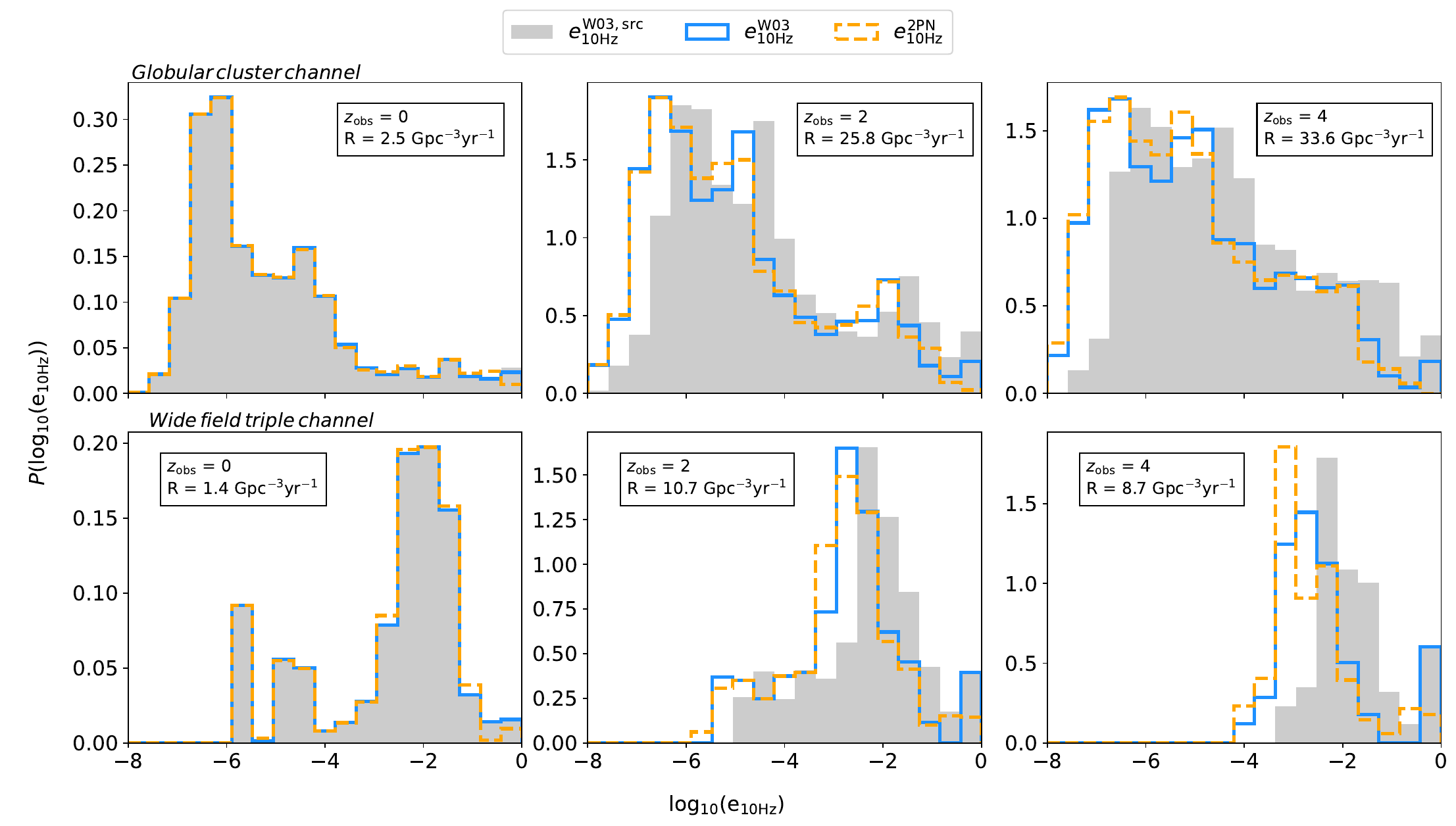}
\caption{The eccentricity distributions of BBH mergers from GCs (upper panel) and wide field triples (lower panel) at three different observed redshifts, $z_{\rm obs}$. We adopt three different definitions for eccentricity extracted at 10 Hz: $e^{\rm W03, src}_{\rm 10Hz}$ and $e^{\rm W03}_{\rm 10Hz}$, eccentricity based on the prescription of \citet{Wen2003}, in the source frame and redshifted to the detector frame, respectively; and $e^{\rm 2PN}_{\rm 10Hz}$, detector-frame eccentricity based on the prescription of \citet{Vijaykumar2024}. For each redshift $z_{\rm obs}$, we show the predicted merger rate density for the specific formation channel.
We have normalised each histogram due to merger rate density at $z_{\rm obs}$.
}
\label{fig:ft_ecc_z}
\end{figure*}

\begin{figure*}
\includegraphics[width=\textwidth]{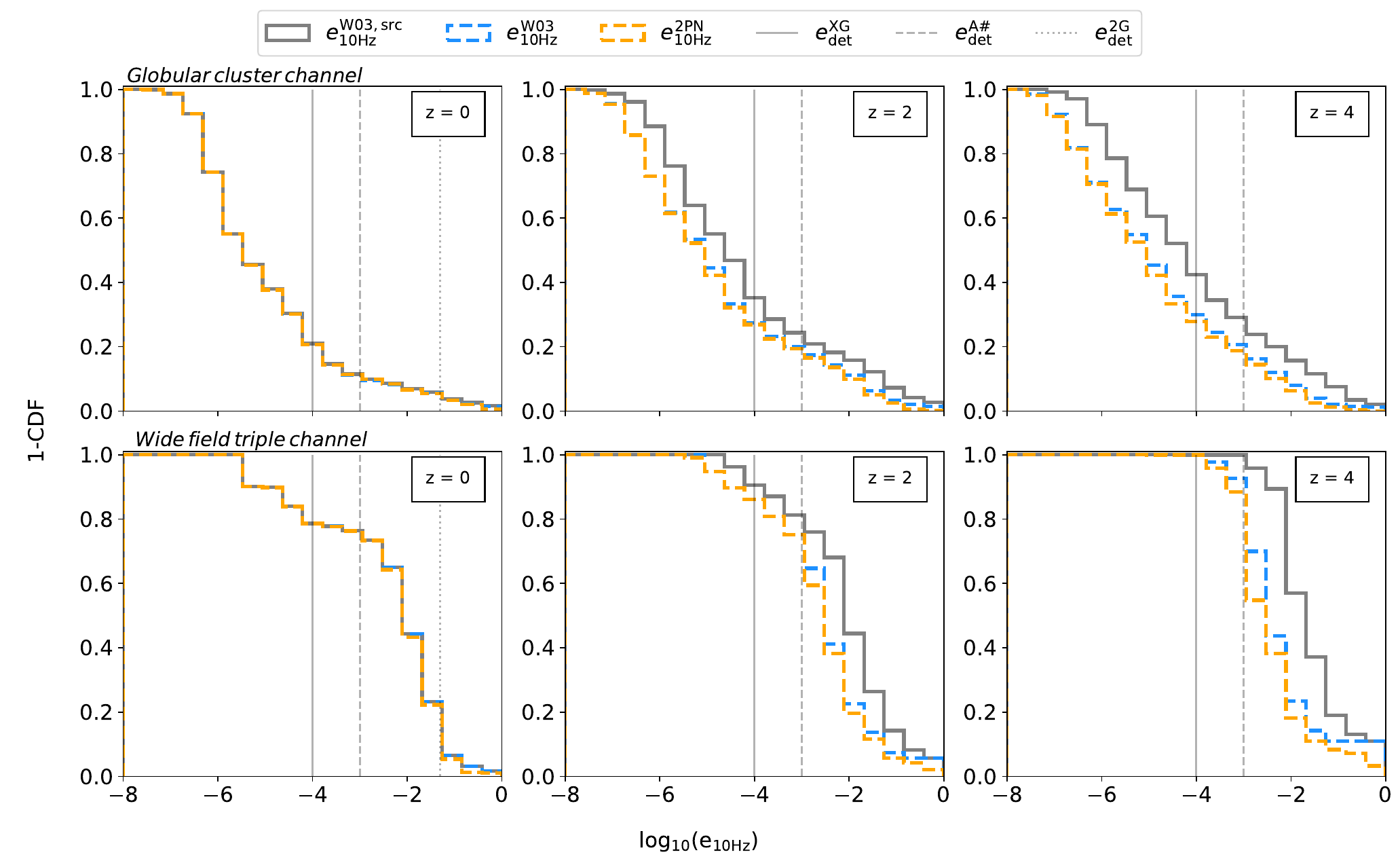}
\caption{
The cumulative eccentricity distribution from GCs (upper panels) and wide field triples (lower panel) at three different redshifts. We adopt three different definitions for eccentricity extracted at 10 Hz; see the caption of Fig. \ref{fig:ft_ecc_z} and Section \ref{sec:detected_eccentricity} for details. We show three different detectable eccentricity thresholds:  $e^{\rm 2G}_{\rm det}$ = 0.05, the eccentricity threshold for 2G detectors, which is only shown for $z_{\rm obs} = 0$; $e^{\rm A\#}_{\rm det} = 10^{-3}$, our optimistic estimated eccentricity threshold for LIGO A$\#$; and finally $e^{\rm XG}_{\rm det} = 10^{-4}$, an optimistic estimate for XG detectors \citep{Lower:Eccentricity:2018}. 
}
\label{fig:ftGC_ecc_redshift_cumul}
\end{figure*}

The convention for extracting $e_\mathrm{10Hz}$ for mergers in simulations of GW progenitor evolution has typically been based on the prescriptions of \citet{Wen2003} (henceforth W03) for extracting eccentricity at a frequency at which the power emitted via GWs peaks (we will refer to this frequency as \textit{peak frequency}).
However, this definition of eccentricity does not align with the one that is actually measured using waveform models. Waveform models have their own internal definitions that can differ substantially between different waveform models, and usually differ from that used in simulations at high eccentricities \citep[$e_\mathrm{10Hz} \gtrsim 0.1$;][henceforth V24]{Shaikh:2023:ecc, Vijaykumar2024}. Furthermore, astrophysical simulations halt the system's evolution once its W03 peak frequency is above the extraction frequency (here $10$~Hz). However, for eccentric systems, the orbit-averaged 22-mode frequency can be much lower than the peak frequency\footnote{Here, the 22-mode frequency refers to the frequency of the (l=2, m=2) mode in the decomposition of the waveform into spin-weighted spherical harmonics (i.e. the quadrupole mode). To a good approximation, the orbit-averaged 22-mode frequency is simply twice the Keplerian orbital frequency. See e.g., \citet{Shaikh:2023:ecc} and V24 for more details}.
This becomes a problem particularly when highly-eccentric sources become bound with a peak frequency greater than the extraction frequency. Simulations will then halt the evolution, and record the measured eccentricity as $\sim 1$. Meanwhile, if the system is further evolved, its eccentricity reduces substantially \textit{before} the 22-mode frequency reaches $10$~Hz, as explained in V24. Additionally, detected signals are redshifted, altering the eccentricity measured at a reference frequency. 

We convert our samples to ``observed'' distributions at $10$~Hz via two methods.
Firstly, we simply redshift the eccentricity distribution output by the simulations (see Sec. 4.1.1 of \citealt{Romero-Shaw:FourEccentricMergers:2022}).
This maintains the W03 definition of the peak frequency at which eccentricity is extracted.
Secondly, we use the conversion scripts available via \citet{aditya_vijaykumar_2025_14674322} to transform simulation-predicted source-frame eccentricities defined at a peak GW frequency of $10$~Hz to detector-frame eccentricities defined at $10$~Hz 22-mode GW frequency, approximating the method described in \citet{Shaikh:2023:ecc}. We note that for a small fraction of sources ($\sim$ 3 per cent), we encountered numerical issues when solving the orbital evolution equations in the conversion script of V24. These issues occur for systems with relatively high detector-frame mass ($M\gtrsim100\,M_{\odot})$ and/or with very small semi-latus rectum $a(1-e^2)$. 
For these systems, we calculated the eccentricity at $10$~Hz 22-mode GW frequency using the analytical formula of \citet{Tucker2021}.

In the following, we differentiate between the two definitions with superscripts: $e_\mathrm{10Hz}^{\rm W03, src}$ for the W03 definition and source-frame predictions, $e_\mathrm{10Hz}^{\rm W03}$ for the latter's detector-frame prediction, and $e_\mathrm{10Hz}^{\rm 2PN}$ for the value obtained via the V24 conversion. 
We find that while the fraction of systems in the $e_\mathrm{10Hz}\sim1$ peak substantially varies with the different definitions, the fraction of systems with detectable eccentricity is relatively robust to these conversions, changing by only a few percent.
The V24 conversion process leads to far fewer binaries with eccentricities near unity.
However, these systems still remain within the detectably-eccentric range, causing relatively small deviations in detectably-eccentric merger rates between different definitions of redshifted eccentricity. 


\subsection{Redshift evolution of eccentricity distributions}

In Fig. \ref{fig:ft_ecc_z}, we show the observed eccentricity distributions and merger rates for the GC channel (upper panel) and for the wide field triple channel (lower panel) for BBH mergers observed at three different redshifts: $z_{\rm obs} = 0,\, 2,\,$ and $4$. 
Both the merger rates and the average eccentricities of GW sources initially increase with increasing redshift, for both formation channels. 



\subsubsection{Globular clusters}
The shape of the eccentricity distribution associated with the GCs changes appreciably with redshift. Notably, the relative merger rate of the lowest-eccentricity peak from ejected binaries that merge outside the cluster decreases, while the higher-eccentricity peaks of the in-cluster and GW capture mergers grow larger with increasing redshift (see Fig. \ref{fig:GCFT_Z}). These trends can be attributed to the distinct delay time distributions associated with the different sub-populations, while the effects of metallicity remain small (see Section \ref{sec:Ztrend}). On average, the delay times associated with ejected mergers are longer than those of the other two sub-populations; the delay-time distribution of ejected mergers peaks around $\sim10\,\rm{Gyr}$, while those of the other sub-populations peak around $\sim1\,\rm{Gyr}$. 

The long delay times associated with the ejected systems lead to an increased merger rate in the local Universe and decreased rate at higher redshifts, i.e. $z_{\rm obs}\gtrsim2$. The opposite trend can be attributed to the shorter delay times of the in-cluster and GW capture mergers.  
This behaviour can be understood in the context of the cosmic star formation model in GCs from \citet{ElBadry2019}, which predicts which predicts that the majority of GC star formation occurred at high redshifts, with approximately 90 per cent of stars forming between $2 \lesssim z \lesssim 8$. Thus, the majority of ejected merging BBHs, that have long delay times, merge at near Universe. On the other hand, GW sources from the in-cluster and GW capture channels typically merger at higher redshifts due to their short delay times.

 \subsubsection{Wide field triples}
 The change in the eccentricity distribution of GW sources from wide field triples is driven by a complex interplay between delay times, the metallicity-dependent SFR, and the metallicity-dependent eccentricity distributions associated with this formation path. The merger rates, as well as the relative contributions associated with the peaks near $e^{\rm W03, src}_{\rm 10Hz} \sim 1$ and $e^{\rm W03, src}_{\rm 10Hz} \sim 0.01$, both increase with increasing redshift up to $z_{\rm obs}\sim4$. This occurs because systems associated with these two peaks form more efficiently in low-metallicity environments (i.e. $Z\lesssim 0.005$, see lower panel of Fig. \ref{fig:GCFT_Z}) and have relatively short delay times. Therefore, their merger rates closely follow the low-metallicity star formation rate, which starts to dominate the total star formation at higher redshifts ($z_{\rm form} \gtrsim 0.8$-$2$). Additionally, we see that there is an appreciable contribution of GW sources from triples that never transition to non-secular dynamical regime in the local Universe but not at higher redshifts. As shown in Fig. \ref{fig:GCFT_Z}, these systems only form efficiently in relatively low-metallicity systems.
Why these GW sources are only significant in the local Universe, can be understood by the long delay times associated with this sub-channel, which are typically in the order of $10\,\mathrm{Gyr}$. While most of the low-metallicity (i.e $Z\lesssim0.002$) star formation occurs at higher redshifts, 
the long delay times of the purely secularly evolving GW progenitors imply that most of them merge in the near Universe.
The absence of these sources at higher redshifts means that at $z_{\rm obs} \gtrsim 4$, essentially all GW sources from the wide field triple channel have eccentricities of $e^{\rm W03, src}_{\rm 10Hz} \gtrsim 10^{-4}$, which is above the lowest detectable eccentricity for XG detectors.

\begin{figure*}
\includegraphics[width=\textwidth]{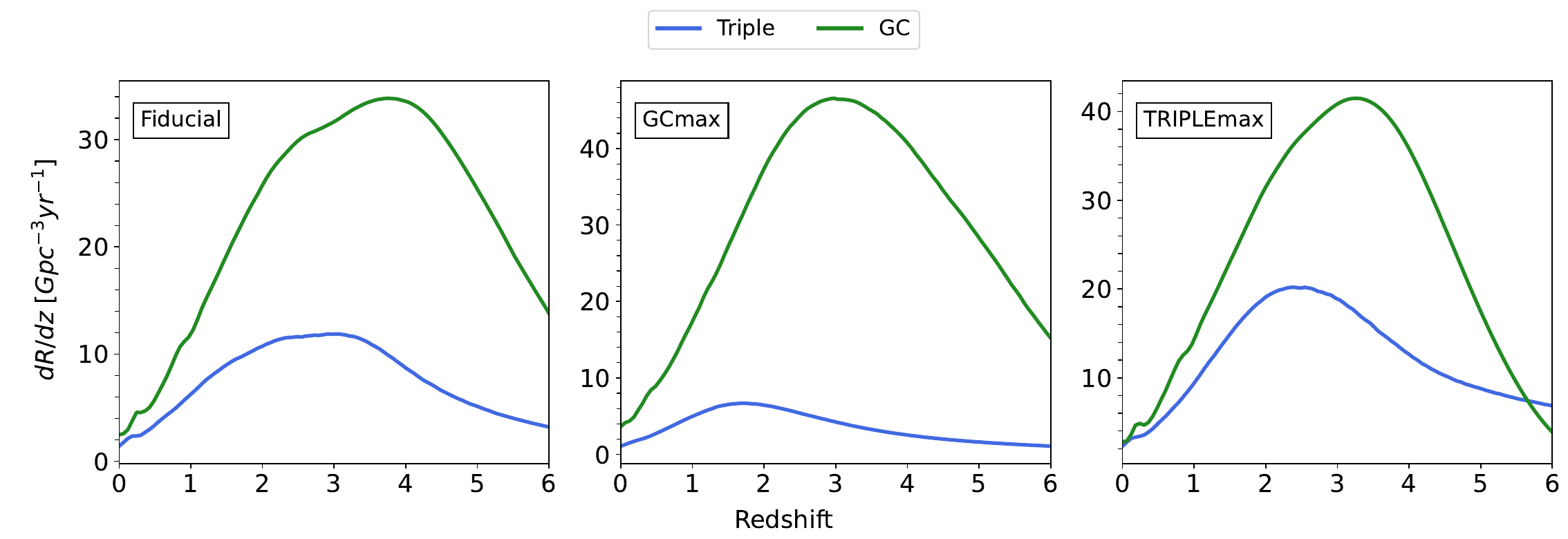}
\caption{The total merger rate of BBHs from wide field triples (blue) and GCs (green) as a function of redshift for the three different model variations considered in this paper (see Table \ref{tab:model_variations}). This includes all mergers, not just those that are detectably eccentric. 
}
\label{fig:Rtotals}
\end{figure*}


\subsubsection{Detected eccentricities in XG detectors}
We proceed to discuss the distributions of detector-frame eccentricities, which provide valuable insights into the prospects of detecting eccentric GW mergers with XG ground-based GW detectors. Fig. \ref{fig:ft_ecc_z} illustrates that detector-frame eccentricities are lower than their corresponding source-frame values due to redshifting. We note that for many systems in the highest-eccentricity bin, $e^{\rm W03, src}_{\rm 10Hz}$ is sufficiently high such that the observed $e^{\rm W03}_{\rm 10Hz}$ remains in that most extremal bin, despite having been evolved from $10$ Hz to $10 (1 + z)$~Hz. 

In Fig. \ref{fig:ft_ecc_z}, we also show the distribution of $e^{\rm 2PN}_{\rm 10Hz}$ detector-frame frame eccentricity calculated according to the prescription of \citet{Vijaykumar2024}. Using $e^{\rm 2PN}_{\rm 10Hz}$ likely provides a more realistic representation of the eccentricity distribution that could be inferred from GW observations, as $e^{\rm 2PN}_{\rm 10Hz}$ is a proxy for the mechanism of \citet{Shaikh:2023:ecc}, which allows eccentricity to be measured directly from the waveform as long as it has enough cycles in-band. The differences between $e^{\rm 2PN}_{\rm 10Hz}$ and $e^{\rm W03}_{\rm 10Hz}$ are negligible below $e^{\rm W03}_{\rm 10Hz}\approx0.2$, while above, $e^{\rm 2PN}_{\rm 10Hz}$ is significantly lower than the corresponding $e^{\rm W03}_{\rm 10Hz}$ value. This contributes to the lack of a peak near $e_{\rm 10Hz}\sim 1$ in the distribution of $e^{\rm 2PN}_{\rm 10Hz}$. 

The minimum detectable eccentricity with XG ground-based GW detectors, such as ET or CE, is estimated to $e_{\rm 10 Hz} \approx 10^{-4}$-$10^{-3}$ at $10$~Hz, with the former obtained with an optimistic overlap calculation \citep{Lower:Eccentricity:2018} and the latter with a Bayesian inference study \citep{Saini:2024:ETsensitivity}, although note that the sensitivity of GW detectors to eccentricity is mass-dependent \citep[see Appendix C of][]{Romero-Shaw:2021:two-ecc}.
We calculate, using the optimistic overlap method used in \citet{Lower:Eccentricity:2018} and a GW150914-like system, that for LIGO A$\#$ the minimum detectable eccentricity is $e_{\rm 10 Hz} \gtrsim 10^{-3}$. This is undoubtedly an underestimate, but a full Bayesian inference study is outside of the scope of this paper, and we relegate that to future work. To determine the fraction of GW sources that have sufficiently high eccentricities to be detected by different current and future ground-based detectors ($f_{\rm e, det}$), we convert the distributions shown in Fig. \ref{fig:ft_ecc_z} to cumulative distributions in Fig. \ref{fig:ftGC_ecc_redshift_cumul}. In the same figure, we also indicate different detection threshold eccentricities : (i) for 2G detectors, $e^{\rm 2G}_{\rm det} = 0.05$, only shown for $z_{\rm obs} = 0$, (ii) for LIGO A$\#$, $e^{\rm A\#}_{\rm det} = 10^{-3}$ and (iii) for XG detectors, assuming the more optimistic value of $e^{\rm XG}_{\rm det} = 10^{-4}$. We note that according to \citet{Saini:2024:ETsensitivity}, $e^{\rm XG}_{\rm det} = 10^{-3}$, which is the same value as our optimistic estimate for $e^{\rm A\#}_{\rm det}$.

The upper panels of Fig. \ref{fig:ftGC_ecc_redshift_cumul} indicate that current detectors are sensitive to the eccentricities of $\approx$ 5 per cent of GW sources from the GC channel, in agreement with previous studies, e.g., \citet{Zevin:EccentricImplications:2021}. This percentage increases up to 20 per cent for XG detectors at $z_{\rm obs} = 0$. For the GC channel, $f_{\rm e, det}$ moderately increases with redshift, reaching 28 per cent at $z_{\rm obs} = 2$ and 30 per cent at $z_{\rm obs} = 4$. The fraction $f_{\rm e, det}$ decreases only slightly, by $\mathcal{O}(1)$ per cent, when the eccentricity definition of \citet{Vijaykumar2024} is adopted instead of the \citet{Wen2003} prescription.

As the lower panels of Fig. \ref{fig:ftGC_ecc_redshift_cumul} show, $f_{\rm e, det}$ is considerably higher for wide field triples than for the GC channel.
About 10-20 per cent of GW sources from field triples have detectable eccentricity for 2G detectors at $z=0$. 
XG detectors will be able to detect the eccentricity of the majority of GW sources from wide field triples; $f_{\rm e, det}$ is about 80-90 per cent for $z_{\rm obs}=0$-$2$ and 100 per cent for $z_{\rm obs} = 4$. Since the eccentricities associated with this channel are distributed over a relatively small range, $f_{\rm e, det}$ is sensitively dependent on the threshold eccentricity of the detector. For example, if the threshold eccentricity is just below $e = 10^{-3}$, then $f_{\rm e, det}$ decreases to 60 per cent and 50 per cent for $z_{\rm obs}=2$ and $z_{\rm obs} = 4$, respectively.

\subsection{Redshift evolution of eccentric merger rates}
\label{sec:Rtrend}
In the left panel of Fig. \ref{fig:Rtotals}, we show the merger rate density as a function of redshift for the GC and wide field triple channel for all eccentricities. The total merger rate is dominated by the GC channel across all redshifts. At $z\sim0$, the merger rate of the channels are comparable ($\sim 1.4-2.5$ $\rm{Gpc^{-3}}\rm{yr}^{-1}$), while at $z\sim2$-$3$, the merger rate associated with the GC channel is about 3 times larger.

The predicted merger rate density of the wide field triple channel in the nearby Universe is in broad agreement with the results of \citet{Antonini2017}. Specifically, their models that adopt the natal kick velocity prescription of \citet{Hobbs2005} yield a merger rate of $R\sim0.5\,\rm{Gpc^{-3}yr^{-1}}$, which is approximately a factor three lower than our estimate. This discrepancy can likely be attributed to three key differences between the models. First, \cite{Antonini2017} only considers $Z = 0.02$ and they do not include lower metallicities in their simulations. Second, they adopt a different kick velocity prescription. Third, they assume a redshift-independent constant star-formation rate.

The evolution of the merger rate density over redshift from the GC channel shows a good agreement with the model of \citet{RodriguezLoeb2018}, which assumes a cut-off mass of $M_{*} = 10^{6}$ in the CIMF (see equation \ref{eq:CIMF}); with both models predicting a peak merger rate density close to $R \sim 30,\mathrm{Gpc^{-3}yr^{-1}}$ (see their fig. 2). In contrast, when no cut-off mass is assumed, the \citet{RodriguezLoeb2018} model predicts a merger rate density roughly three times higher than ours. Similarly, \citet{Kremer2020} finds a merger rate density that is about five times greater than our estimate. The difference can be explained by the differences in the adopted CIMF and in the assumed present-day cluster density, which is approximately 1.5 times higher in their model. Since our models, as well as those of \citet{RodriguezLoeb2018} and \citet{Kremer2020}, use \texttt{CMC} with the GC formation model of \citet{ElBadry2019}, we can see that variations in the assumed initial conditions of clusters can significantly impact the predicted GW merger rate density. We explore this further in Section \ref{sec:uncertainties}.

If one focuses on only GW sources with detectable eccentricities, the GC channel is no longer the dominant formation path.
In the upper row of Fig. \ref{fig:model_variatons}, we present the merger rate density evolution over redshift for GW sources with detectable eccentricity, using our three detectable eccentricity threshold values and three eccentricity definitions. The merger rates of GW sources with detectable detector-frame eccentricities reach about $7$ $\rm{Gpc^{-3}yr^{-1}}$ and $9$ $\rm{Gpc^{-3}yr^{-1}}$ at their peak for the GC and the wide field triple channel, respectively, when considering an detectable eccentricity threshold of $10^{-3}$. 
In the local Universe ($z\lesssim0.5$), wide field triples moderately dominate the eccentric merger rate for all eccentricity thresholds and definitions. The eccentric merger rate for both threshold eccentricities continues to be dominated by triples up to $z\lesssim 4$, while for a 2G-detector threshold of $e_{10 \rm Hz} > 0.05$ the eccentric merger rate for triples and GCs is similar across a wide range of redshifts. 
We note that our predicted source-frame eccentric GW merger rate for the GC channel, assuming a detector threshold of $e_{10 \rm Hz} > 0.05$, is in reasonable agreement with the predictions shown for the in Fig. 10 of \citet{Rodriguez2018:formationmassesmergerrates}. According to their results, the merger rate of GW sources from the GC channel with $e_{10 \rm Hz} > 0.05$ reaches a maximum of $\sim$6 $\rm{Gpc^{-3}yr^{-1}}$, which is about a factor of 2.5 larger than than our model predictions. This difference can be entirely due to the different CIMF assumed in \citet{Rodriguez2018:formationmassesmergerrates}. 

Our main prediction is that eccentric BBH mergers observed with GWs are moderately dominated by the wide field triple channel over the GC channel in the local Universe for current detectors, as well as at redshifts $z_{\rm obs}\lesssim4$ for future detectors. These findings hold regardless of which definition of eccentricity is adopted. This highlights that eccentric GW mergers cannot be solely associated with formation channels of dense environments, as the majority of these sources could be produced in field stars at a wide range of redshifts.

\subsection{Model variations}
\label{sec:uncertainties}
\begin{figure*}
\includegraphics[width=\textwidth]{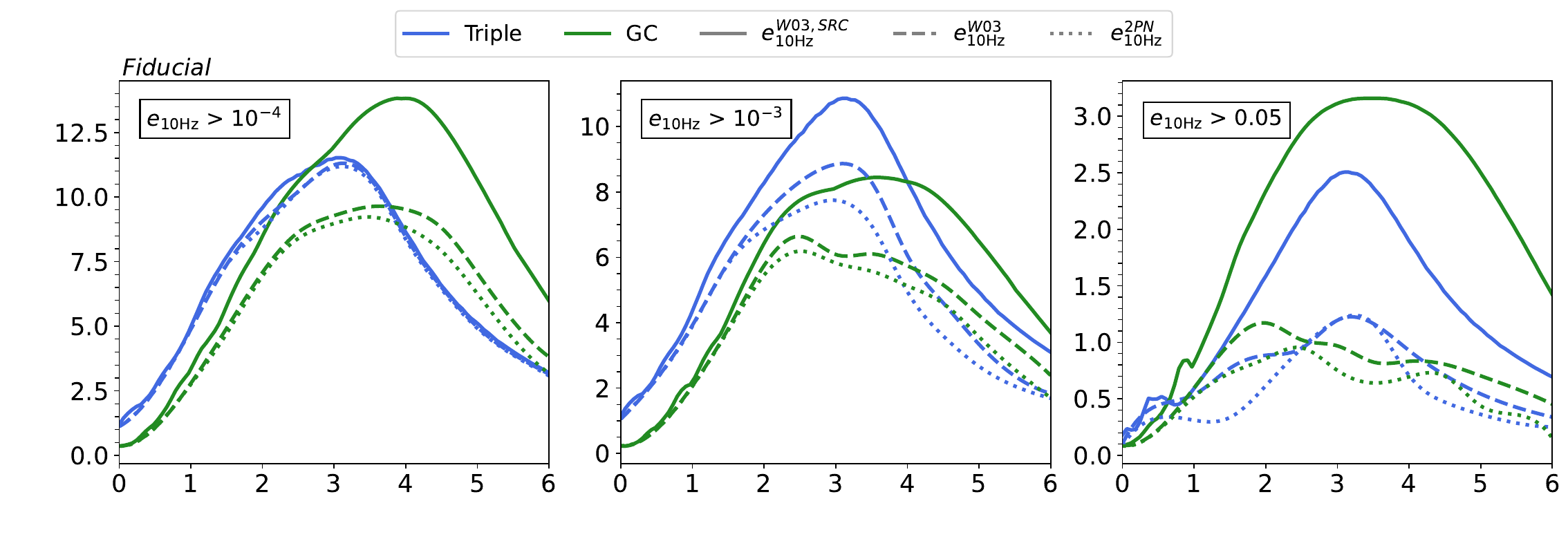}\hfill
\includegraphics[width=\textwidth]{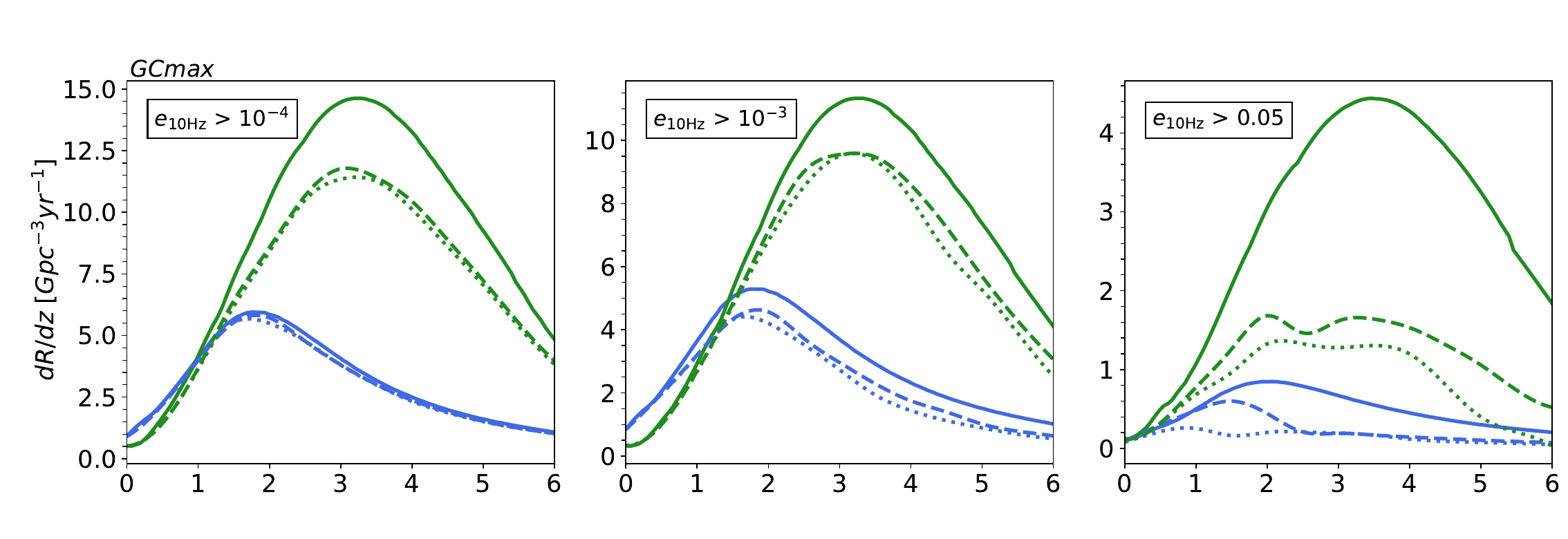}\hfill
\includegraphics[width=\textwidth]{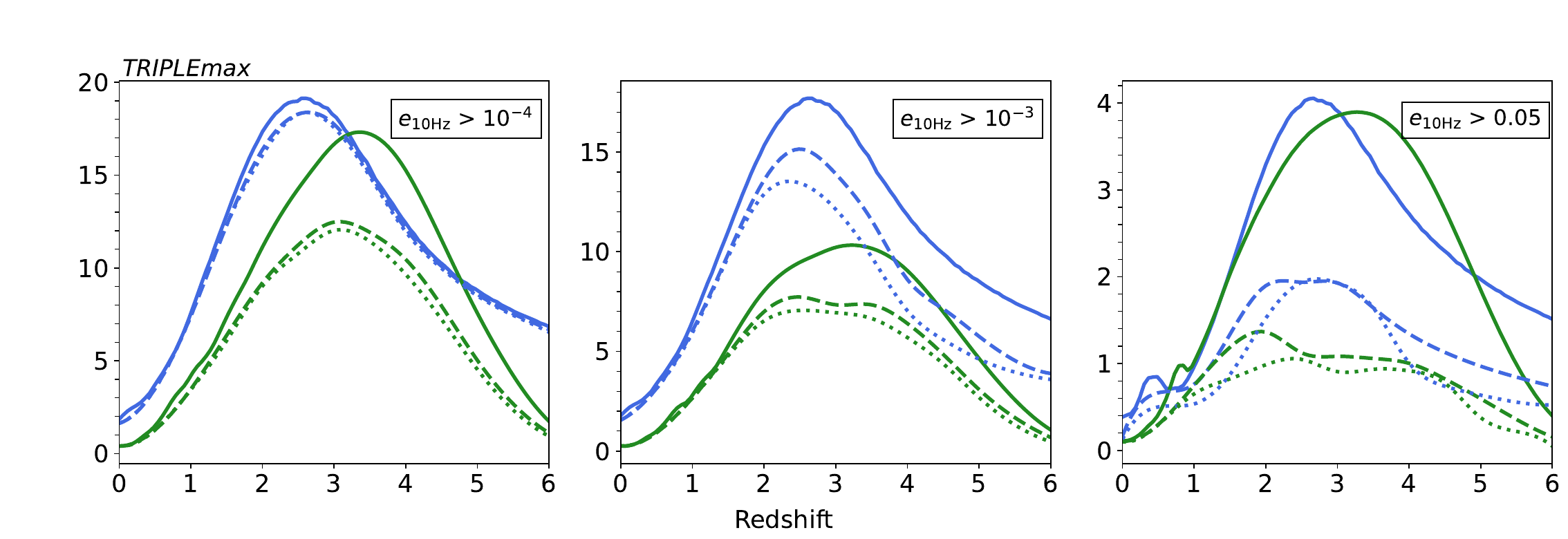}
  \caption{We highlight the robustness of our main results to uncertainties by showing  the detectably-eccentric merger rates assuming thresholds of 
  $e_{\rm10Hz}> 0.05$ (right panel),
  $e_{\rm10Hz}> 10^{-4}$ (left panel) and $e_{\rm10Hz}> 10^{-3}$ (middle panel) for different model variations: fiducial (upper panel), \textit{GCmax} (middle panel), \textit{TRIPLEmax} (lower panel). For each panel, we denote different eccentricity definitions with different linestyles: $e^{\rm W03, src}_{\rm 10Hz}$, source-frame eccentricity based on the prescription of \citet{Wen2003} (solid), $e^{\rm W03}_{\rm 10Hz}$, detector-frame eccentricity based on the prescription of \citet{Wen2003} (dashed, and $e^{\rm 2PN}_{\rm 10Hz}$, detector-frame eccentricity based on the prescription of \citet{Vijaykumar2024} (dotted). 
  }
  \label{fig:model_variatons}
\end{figure*}

In our fiducial model, the merger rate of GW sources from wide field triples with detectable eccentricities in the detector frame is at most twice that of the GC channel (Fig. \ref{fig:model_variatons}). Given this moderate difference and the numerous uncertainties in our models (detailed in Section \ref{subsec:weighing_scheme}), we must assess the robustness of the qualitative result that triples dominate the eccentric merger rate. We compare the predicted merger rate density distribution over redshift for three different model variations in Fig. \ref{fig:model_variatons}. Detailed descriptions for these model variations are given in Section \ref{subsec:weighing_scheme} and summarised in Table \ref{tab:model_variations}.


\begin{figure*}
\includegraphics[width=\textwidth]{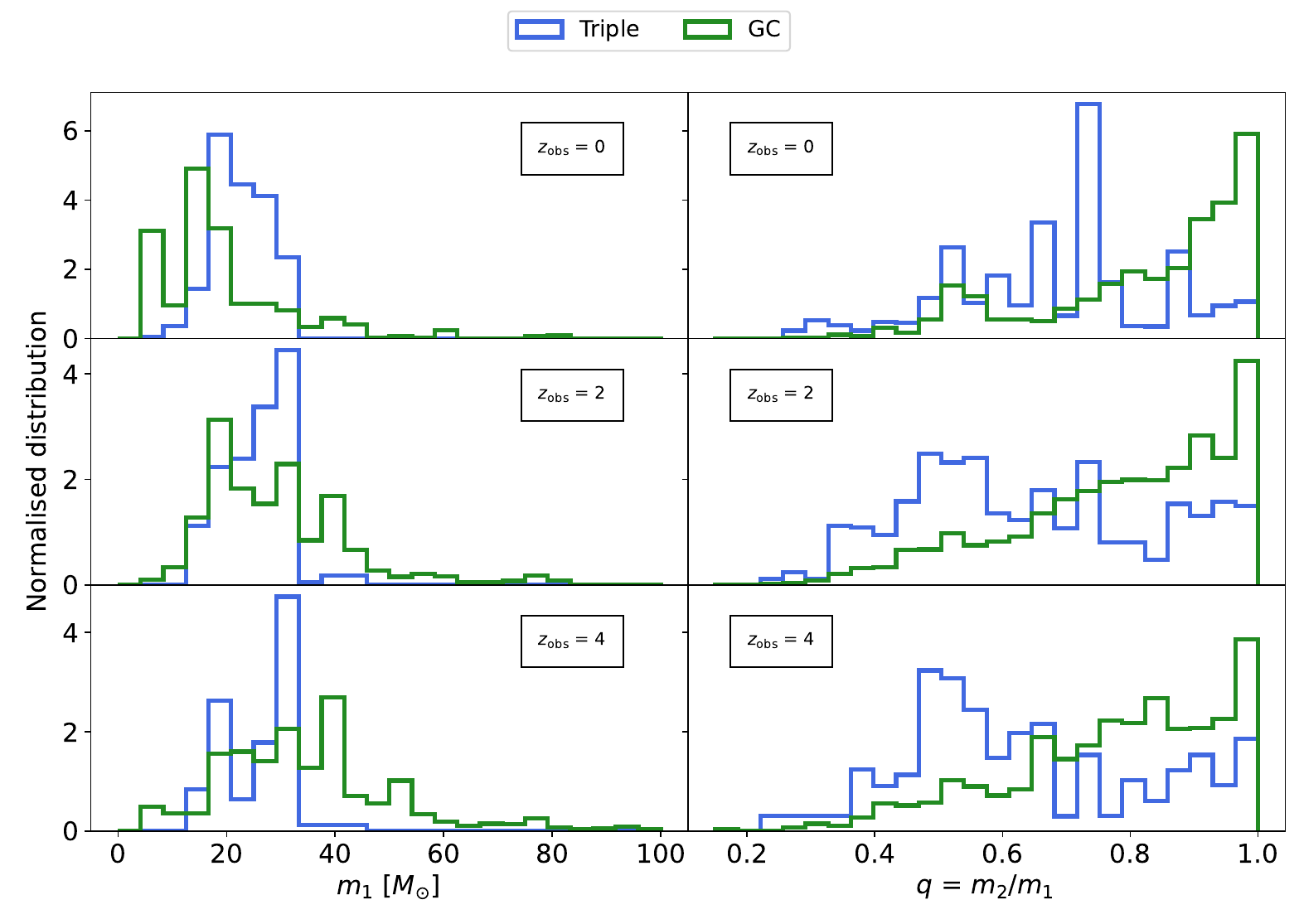}
\caption{The primary mass (left) and mass ratio (right) distributions of the merging binary black holes for the GC channel (green) and for the wide field triple channel (blue) for BBHs merging at three different redshifts, i.e. $z_{\rm obs} = 0$, $2$, $4$. All histograms are independently normalised to arbitrary units. Therefore the difference between the area under each histogram in a specific panel does not reflect the difference in merger rate density. 
} 
\label{fig:primary_and_ratio}
\end{figure*} 

In the \textit{GCmax} model, we choose uncertain parameters related to GC formation that favor the efficiency of GW source formation in GCs over field triples (see Table \ref{tab:model_variations} and disucssion in section \ref{subsec:model_variations}). 
As shown in Fig. \ref{fig:model_variatons}, in the \textit{GCmax} model the merger rate of detectably eccentric BBHs from GC channel is only moderately increased compared to the fiducial model. For a given GC formation model, increasing the cut-off mass $M_*$ is known to significantly increase the merger rate, as GW sources form more efficiently in more massive and denser clusters \citep[see, e.g.,][]{RodriguezLoeb2018}. In our models, however, we derive the initial GC mass density $\rho_{\rm GC,0}$ from the adopted CIMF instead of treating as a fixed parameter (see equation \ref{eq:Kev} and discussion). This leads to a different normalisation of the cluster formation model of \citet{ElBadry2019}. A CIMF favouring higher mass clusters implies less efficient GC evaporation and therefore a lower $\rho_{\rm GC}$. These two opposing effects ultimately lead to only a moderate increase in the merger rate associated with the GC channel. Meanwhile, the merger rate from the wide field triple channel decreases by a factor of two, relative to the fiducial model. This is due to the adopted higher-metallicity SFR model, which leads to a lower formation efficiency of GW sources in wide triples.


Even in the \textit{GCmax} model, the contribution from the wide field triple channel remains significant. As shown in the middle and left panels of the middle row in Fig. \ref{fig:model_variatons}, the contributions from the two channels are comparable at low redshifts ($z_{\rm obs} \lesssim 1$). At $z_{\rm obs}\approx 2$ around half, and at $z_{\rm obs}\approx 5$ about one-fourth of all eccentric GW mergers originate from field triples in this model. The merger rate density of detectably-eccentric GW sources from wide triples reaches a peak value of 4.5 $\rm{Gpc^{-3}yr^{-1}}$
around $z_{\rm obs} \approx 2$. These results show robustly that GW progenitors formed from field stars may contribute significantly to eccentric mergers and therefore 
eccentric GW mergers cannot be considered a definitive indicator for formation in a dense environment.

The bottom row of Fig. \ref{fig:model_variatons} shows results from the \textit{TRIPLEmax} model, in which we adopt parameters that enhance the formation efficiency of GW sources from wide field triples and suppress that of GCs.
Under this model, the detector frame detectably-eccentric GW mergers are dominated by field triples over the entire redshift range, reaching 15-20 $\rm{Gpc^{-3}yr^{-1}}$ around $z_{\rm obs}=2.5$, depending on the threshold eccentricity definiton. The contribution GCs remains important, with a peak merger rate density of 10-12 $\rm{Gpc^{-3}yr^{-1}}$, around $z_{\rm obs}=3.5$.


We emphasize that there are other important uncertainties which we have not explored in this work. For the GC channel, the most significant among these are the present-day GC number density and the initial densities of clusters. Typical uncertainties in these parameters can lead to a factor of 2–3 variation in the predicted merger rate \citep[see e.g.][]{AntoniniGieles2020}. Uncertainties related to GC evaporation factor $K_{\rm ev}$ are also substantial. \citet{AntoniniGieles2020} derives the posterior distribution for $K_{\rm ev}$ by fitting the present-day GC mass function $N_{\rm CIMF, 0}$ to the catalog of \citet{Harris2010} and finds $K_{\rm ev} = 32.5^{+86.9}_{-17.7}$. The large error bars indiciate that this parameter alone can introduce an order of magnitude uncertainty in the predicted merger rate from the GC channel.

While uncertainties related to stellar evolution is less important for the GC channel \citep[see e.g.][]{AntoniniGieles2020},  they can have a substantial impact on predictions for the wide field triple channel. In particular, results \citet{Antonini2017} shows that uncertainties in the natal kick remain a key factor for this channel, as their predicted merger rate varies by a factor of three depending on the natal kick prescription assumed for black holes.

With their vastly improved sensitivity to high-redshift mergers and lower-frequency sensitivity, XG detectors are ideal instruments for distinguishing redshift-evolving eccentricity distributions. The prediction of a lower-redshift peak of the eccentric merger rate for triples than GCs is robust to our model variations, making this an observable distinguishing feature of the two channels.


 \subsection{Distinguishing the GC and wide field triple channel}
 \label{subsec:distinguishing}
Distinguishing the formation channels of GW sources is crucial for understanding the formation environments of GW progenitors and the processes that drive their evolution. However, the contributions of GC and the wide field triple channel to the eccentric merger rate are comparable across a wide range of redshifts, with different channels dominating at different redshifts (see, e.g., Fig. \ref{fig:model_variatons}). Identifying the formation channel of a single eccentric merger therefore requires inspection of additional 
parameters, while distinguishing different sub-populations of eccentric mergers may be achieved through studying parameter distributions. Here, we discuss 
GW observables that can help differentiate between the two channels studied in this work.

In Fig. \ref{fig:primary_and_ratio}, we show the primary mass and mass ratio distributions associated with the two channels at $z_{\rm obs} = 0,$ $2,$ $4$ \cite[see also related work of][]{Ye2024:bbh_mass_distribution}. There are two important distinguishing features in these distributions. Firstly, the primary mass distribution associated with the GC channel extends beyond $\sim100\,M_{\odot}$, and this high mass tail becomes more prominent with increasing redshift. On the other hand, the maximum mass of the merging binary black holes from the wide field triple channel remains below the uncertain maximum mass limit due to pair-instability supernovae, here assumed to be $\sim45\,M_{\odot}$ \citep[e.g.][]{Fowler1964, Rakavy1967, Barkat1967, Fraley1968, Yoshida2016, Marchant2016, Woosley17, Renzo2020}, irrespective of redshift. Observational evidence for the pair-instability mass gap and the natal spin of BHs is limited \citep[although see][]{Antonini2025}, and theoretical studies reach conflicting conclusions \citep[e.g.,][]{Takahashi2018, Farmer2019A, WoosleyHeger:2021:MassGap, Costa2021, Mehta2022, Farag2022, MillerMiller:2015:natalspins, FullerMa2019, Belczynski2020:evolutionary_roads_leading_to}. Regardless of its value, this limit can be surpassed in dense environments (such as in GCs) via hierarchical mergers, which build up higher-mass BHs by involving merger products in future coalescences \citep[see, e.g.,][and references within]{Gerosa:2021:Review}, or pre-collapse stellar mergers \citep[e.g.,][]{2021MNRAS.507.5132D}; note that the latter are not accounted for in \texttt{CMC}.

The mass ratio distributions associated with the two channels are also distinct. While BBHs from both channels have a wide range of mass ratios, those from the GC channel favour equal ($q\sim1$) mass ratios, and show relatively weak redshift dependence. On the other hand, those from the wide field triple channel are more uniformly distributed compared to the GC channel 
(in broad agreement with \citealt{Su2021} and \citealt{Martinez2022}), with a peak that shifts to increasingly lower mass ratios with increasing redshifts. In particular, at $z_{\rm obs}\sim4$, a large fraction of black hole binaries from the triple channel have mass ratios $0.4 \lesssim q \lesssim 0.6$. 

Spins of BHs have been long recognised as key GW observables to distinguish different formation pathways \citep[see, e.g.,][]{Vitale15, FragioneKocsis2020, Gerosa17, Kalogera:2000:SpinTilt}. The spin parameter that can be inferred with highest precision from GW signals of inspiralling objects is the effective binary spin ($\chi_{\rm eff}$), the mass-weighted average of the projected spins of components onto the angular momentum vector of the binary \citep[e.g.,][]{Ajith2011}. Spins of merging BBHs formed in spherically-symmetric dense environments such as GCs are predicted to have isotropically-distributed tilts \citep[e.g.,][]{Vitale15, Rodriguez2016spin, Gerosa17}, while merging BBHs from wide field triples may form preferentially with spins that lie in the orbital plane of the inner binary, if the merger occurs over several Kozai cycles \citep[e.g. if the octupole term is relatively weak, see e.g.][]{Liu2017, Liu2018, Antonini2018, RodriguezAntonini2018, FragioneKocsis2020}, or may follow a uniform distribution in $\chi_{\rm eff}$, if the merger occurs abruptly, once a very large eccentricity is reached \citep[e.g., if the octuple term is relatively strong, or the triple follows non-secular evolution, see e.g.][]{Liu2018, RodriguezAntonini2018, Stegmann2025}. It is generally expected that GW sources with detectable eccentricities would belong to the latter group \citep[see e.g.][]{RodriguezAntonini2018, Stegmann2025}.

 The distinct spin orientation distributions of the two formation channels considered in this work also leads to distinct $\chi_{\rm eff}$ distributions, particularly for eccentric sources \citep{Stegmann2025}, and therefore can serve as a key feature to distinguish these formation channels. Additionally, the $\chi_{\rm eff}$ distributions associated with these two channels may evolve distinctly with redshift. Future studies predicting the redshift evolution of spins from different channels may be important, as the BBHs in the GWTC-3 catalog \citep{GWTC-3} show evidence of the spin distribution broadening with redshift \citep[][]{Biscoveanu:2022:RedshiftSpins}. 

The distinct shape and redshift evolution of the eccentricity distributions of the two channels can serve as an additional distinguishing feature. The eccentricity distribution associated with wide field triples shows a Gaussian-like peak centred around $e_{\rm 10 Hz}\approx10^{-2}$, while that of the GC channel is approximately uniformly distributed in the range of $10^{-4} \lesssim e_{\rm 10 Hz} \lesssim 10^{-3}$ (see Fig. \ref{fig:ft_ecc_z}). We find that this prediction is robust against uncertainties explored in this paper. On the other hand, a Gaussian-like peak arises in the eccentricity distribution for GCs if we consider GCs with relatively low initial number densities (see, e.g., leftmost panel in Fig. \ref{fig:e_GC_rv}). This hints that the eccentricity distributions of GC channel can also exhibit a peak around $e_{\rm 10 Hz}\approx10^{-2}$ under certain initial virial radii distributions. We note that the eccentricity distribution is comparatively insensitive to assumptions about BH natal spins and the location of the pair-instability mass gap \citep[e.g.,][]{Samsing17, Rodriguez2018:formationmassesmergerrates, Renzo_2021, Hendriks2023}. 


    
\section{Conclusions}
\label{sec:conclusions}


In this work, we have investigated how the eccentricity distributions of GW sources evolve with redshift from two different formation channels: wide field triples and GCs. 

Our most important conclusions are:
\begin{itemize}
\item \textbf{The fraction of mergers with detectable eccentricity at $10$~Hz increases with redshift for both channels}; see, e.g., Fig. \ref{fig:ftGC_ecc_redshift_cumul}. 
\item \textbf{The merger rate of detectably-eccentric GW sources in XG GW detectors increases appreciably with redshift out to $z \approx 4$}; see, e.g., Fig. \ref{fig:model_variatons}.  
\item{ 
\textbf{The majority of detectably-eccentric GW mergers could be produced by the wide field triple channel.}} Our fiducial model predicts that detectably-eccentric mergers are dominated by the wide field triple channel across a significant range of redshift; $z_{\rm obs} \sim 0$-$4$ (see, e.g., Fig. \ref{fig:model_variatons}). Although the total merger rate from the GC channel exceeds that of the wide-field triple channel \citep[see also][]{Antonini2017, RodriguezAntonini2018}, a larger fraction of GW sources from wide field triples have XG-detectable eccentricities (i.e. $e_{\rm 10 Hz} \geq$ $10^{-3}$-$10^{-4}$) when compared to those from GCs (see, e.g., Fig. \ref{fig:ftGC_ecc_redshift_cumul}).  We also note that the vast majority of GW sources from this channel evolve via non-secular evolution \citep[in agreement with][]{Antonini2017}.
\item \textbf{The significant contribution of the wide field triple channel to detectably-eccentric GW mergers is robust to model uncertainties.} We find that the contribution of wide field triples to detectably eccentric GW mergers remains significant when we vary initial cluster properties and metallicity-dependent SFR models. 
Even when we assume conditions that disfavour the formation efficiency of the wide field triple channel, we find that half (one-fourth) of all detectably-eccentric GW mergers originate from the wide field triple channel at $z_{\rm obs}\approx2$ ($z_{\rm obs}\approx5$) -  see middle row in Fig. \ref{fig:model_variatons}. These results challenge the commonly-held view that eccentric GW mergers necessarily imply a formation channel linked to dense environments (see, e.g., \citealt{OLeary2009, Breivik2016, Samsing17, Zevin:EccentricImplications:2021}, but also see, e.g., \citealt{Antonini2017, RodriguezAntonini2018} for earlier works discussing eccentric mergers from field stars). 
\item \textbf{The eccentricity distributions from the GC and wide field triple channels evolve distinctly with redshift.} This behaviour can therefore be considered a key distinguisher of these two formation channels. In particular, the eccentricity distribution associated with wide field triples shows a Gaussian-like peak centered around $e_{\rm 10 Hz}\approx10^{-2}$, while that of the GC channel is approximately uniformly distributed in the range of $10^{-4} \lesssim e_{\rm 10 Hz} \lesssim 10^{-3}$ (see Fig. \ref{fig:ft_ecc_z}). We find that this prediction is robust against uncertainties explored in this paper. 
\item \textbf{The eccentricity distribution and formation efficiency of mergers in GCs depend weakly on metallicity, in contrast with mergers in wide field triples.} Our results suggest that the merger rate density in GCs should closely follow the star formation history in GCs, offset due to the delay time of GW sources. Consequently, GWs from mergers in GCs can serve as unique tool for constraining star formation in those environments, and models of GC formation in general. However, our results clearly demonstrate that multiple channels can contribute to detectably-eccentric mergers at rates that may compete with those from GCs at a range of redshifts; studies into probing GCs using GWs \citep[e.g.,][]{Romero-Shaw:2021:GCs, FishbachFragione:2023:GCs} should take this into consideration.
\end{itemize}

We predict that the merger rate of detectably-eccentric GW sources will increase with redshift out to $z \approx 3$, while the eccentricity distributions of the two channels studied in this work remain distinct at all redshifts. This implies that XG detectors will enable the mapping of the demographics of these channels via the evolution of the eccentric merger rate with redshift, and distinguish these mergers from GW sources formed through isolated binary evolution channels. 
As noted in Section \ref{sec:intro}, GCs and wide field triples are not the only sources of detectably eccentric mergers. Highly eccentric merging BBHs may form in AGN disks \citep{Samsing22}, or in nuclear clusters \citep{Fragione19, GondanKocsis2021} or may originate from field triples perturbed by flybys \citep{MichaelyPerets2020} or galactic tides \citep{grishin22, steg24}. We have here focused on GCs and wide triples as these systems are relatively well-studied and existing simulations are available for us to access. Future work on the redshift evolution of eccentricity and their associated formation environments would enable us to distinguish between more formation scenarios.

Finally, we note that there has been recent evidence of eccentricity in a neutron star-black hole binary (NSBH) event, GW200105 \citep{2025arXiv250315393M}, and that this and all other NSBHs detected by the LVK are consistent with coming from field triples \citep{2025arXiv250609121S}. While we have focussed exclusively on BBH mergers in this work, a high rate of NSBH mergers from field triples could indicate a high rate of BBH mergers originating from the same environment. We will investigate the redshift-evolving eccentricity distribution of NSBH mergers from field triples in comparison to other environments in future work.

\section*{Acknowledgements}
We thank Carl Rodriguez for sharing the fitting formula for the globular cluster formation model. We thank Ilya Mandel, Giulia Fumagalli, Srija Chakraborty and Hyung Mok Lee for useful discussions. IMR-S acknowledges support received from the Herchel Smith Postdoctoral Fellowship Fund and the Science and Technology Facilities Council grant number ST/Y001990/1. ST acknowledges support from the Netherlands Research Council NWO (VIDI 203.061 grant). AV acknowledges support from the Natural Sciences and Engineering Research Council of
Canada (NSERC) (funding reference number 568580).
MZ gratefully acknowledges funding from the Brinson Foundation in support of astrophysics research at the Adler Planetarium. EG acknowledges support from the ARC Discovery Program DP240103174 (PI: Heger). 
FA is supported by the UK’s Science and Technology Facilities Council
grant ST/V005618/1.
Most numerical computations were carried out on the PC cluster at the Center for Computational Astrophysics, National Astronomical Observatory of Japan. The authors are also grateful for computing resources maintained by the LIGO Laboratory, funded by National Science Foundation Grants PHY-0757058 and PHY-0823459.

\section*{Data Availability}

The data used in this work will be made available upon reasonable request to the corresponding author.


\bibliographystyle{mnras}
\bibliography{example} 




\appendix


\section{Impact of different supernova and natal kick models in \texttt{TRES} and \texttt{CMC}}
\label{appendix:impact_sn_kick}

To estimate the impact of the different supernova and natal kick models used in \texttt{TRES} and \texttt{CMC}, we performed an additional simulation with \texttt{TRES}, adopting the same models as in \texttt{CMC}; the \textit{Rapid} supernova model of \citet{Fryer_2012} and the natal kick prescription given by equation \ref{eq:cmc_kick}. We set the metallicity $Z = 0.0002$, because the vast majority of GW sources from the wide field triple form most efficiently in low metallicity environments (see section \ref{sec:Ztrend}). We find that the formation rate of non-secular BH triples, which is closely tracks the GW merger rate, is about 30 per cent higher with the default models used in this paper than with the supernova and natal kick models adopted from \texttt{CMC}.

This relatively small difference can be attributed to the low metallicity of the majority of the GW sources from the wide field triple channel. At $Z\lesssim0.002$ the \textit{Rapid} and \textit{Delayed} models predict a very similar BH mass spectra. More importantly, at such low metallicities, massive stars with $M_{\rm ZAMS}\gtrsim32\,M_{\odot}$ form BHs via direct collapse, meaning no kicks are imparted to the BH, regardless of which kick model is applied (see equations \ref{eq:kick} and \ref{eq:cmc_kick}). In particular at $Z\lesssim0.002$, in about 47 per cent of our simulated triple systems, none of the stars recieve natal kicks in either of the SN models. While the different BH formation models used in the two codes remain an important caveat, we do not expect that these change our conclusions significantly.


\section{Additional Figures}
\label{appendix:changing-GC-e-with-rv}

\begin{figure*}
\includegraphics[width=0.6\textwidth]{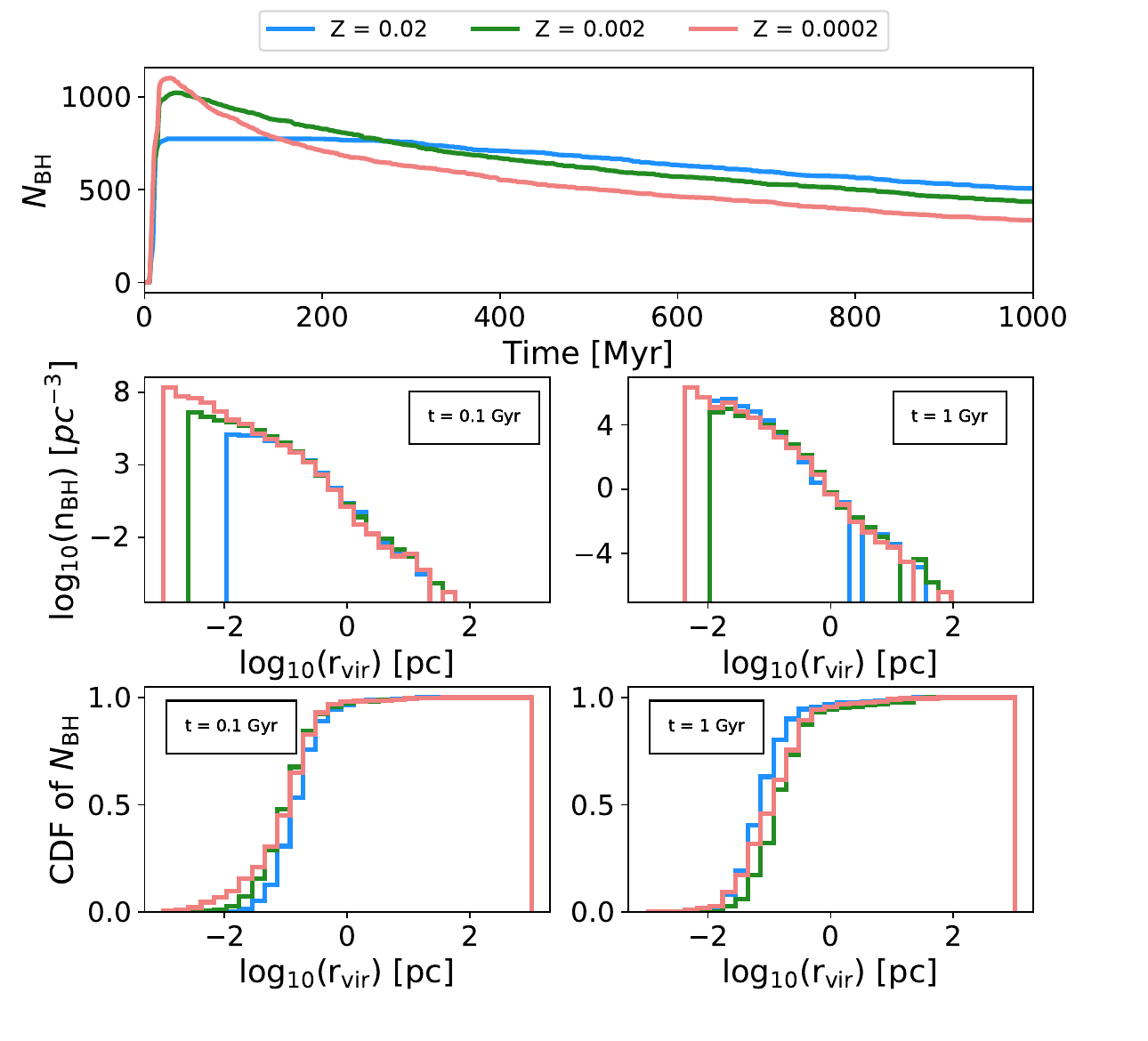}
\caption{Upper panel: the number of BHs in GCs with a virial radius of $r_{\rm vir} = 0.5\,\rm{pc}$ and $N = 8\times10^5$ initial particles and a galactrocentric radius of $R_{\rm GC} = 8\,\rm{kpc}$ and for three different metallcities, Z = 0.02 (blue), 0.002 (green) and 0.0002 (red). See Fig. 7 in \citet{Kremer2020} for a similar figure. Middle panel: the number density of BHs in the same GCs at $t=100\,\rm{Myr}$ (lower left panel) and $t=1000\,\rm{Myr}$. Lower panel: the cumulative distribution of BHs as a function of virial radius at $t=100\,\rm{Myr}$ and $t=1000\,\rm{Myr}$.}
\label{fig:density_profile_gc}
\end{figure*}

In Figure \ref{fig:density_profile_gc}, we explore how GC metallicity influences the number density profile of the GC, and in turn, the eccentricity distribution of its GW mergers. 
Specifically, in this figure, we show the number of BHs as a function of time in three GCs with identical initial parameters, but different metallicities: $Z = 0.02$, $Z = 0.002$ and $Z = 0.0002$. Additionally, we present the number density profiles and the cumulative distributions of BHs within these three cluster models at $t = 100\,\rm{Myr}$ and $t = 1\,\rm{Gyr}$. Shortly after all the BHs form ($t \approx100\,\rm{Myr}$), the lowest metallicity GC retains more BHs due to lower natal kicks of BHs \citep[see also][]{Kremer2020}.  Consequently, BHs in the lowest metallicity GC sink closer to the center, creating a very high number density region within $r \lesssim 0.02\,\rm{pc}$, containing roughly 15 per cent of all BHs in this cluster. This high density region is absent in the solar metallicity GC, leading to initially slightly different number density profiles. However, as high number density also leads to efficient ejection of BHs, this region gets depleted quickly leading to similar number density profiles for all three GC models by $t = 1\,\rm{Gyr}$. 

\begin{figure*}
\includegraphics[width=0.6\textwidth]{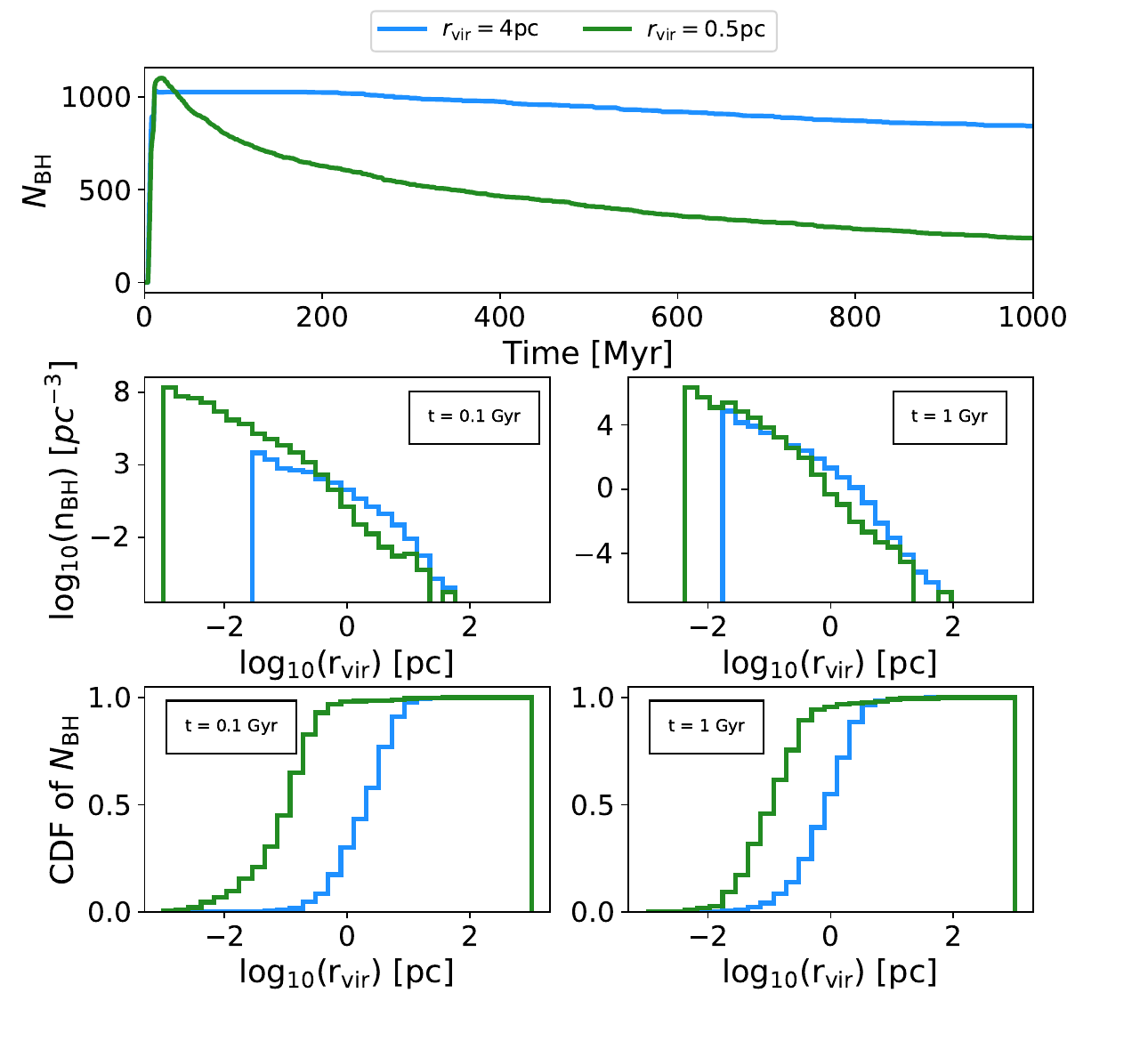}
\caption{The same as Fig. \ref{fig:density_profile_gc} but here we vary the initial virial radii instead of metallicity. All models have a metallicity of Z = 0.0002, initial particles of $N = 8\times10^5$ and a galactrocentric radius of $R_{\rm GC} = 8\,\rm{kpc}$.}
\label{fig:density_profile_rvcomp}
\end{figure*}

\begin{figure*}
\includegraphics[width=\textwidth]{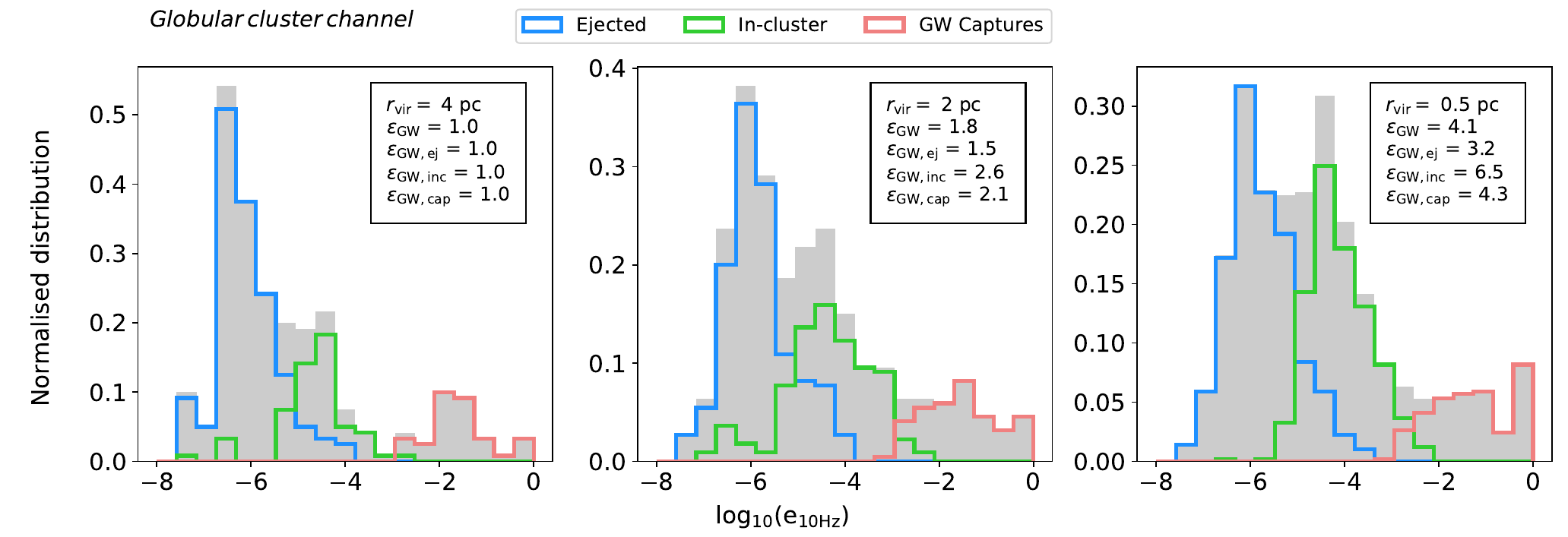}
\caption{The eccentricity distribution of GW sources from the GC channel for three different GC models, with varying virial radii; $r_{\rm vir} = 4\,\rm{pc},\,2\rm{pc},0.5\,\rm{pc}$. All three GC models have Z = 0.0002, an initial particles of $N = 8\times10^5$ and a galactrocentric radius of $R_{\rm GC} = 8\,\rm{kpc}$. In each panel, we show the formation efficiency of all GW sources ($\epsilon_{\rm GW}$) and of all three channels ($\epsilon_{\rm GW, ej}$ for ejected, $\epsilon_{\rm GW, inc}$ for in-cluster and $\epsilon_{\rm GW, cap}$ for GW capture channel) expressed as a fraction of the same formation efficiency for the $r_{\rm vir}$ model. This highlights how the formation efficiency of different subchannels changes with changing number density.}
\label{fig:e_GC_rv}
\end{figure*}

As a comparison, we also include Fig. \ref{fig:density_profile_rvcomp}, in which we compare identical cluster models except for their initial virial radii ($r_{\rm vir} = 4\,\rm{pc}$ and $r_{\rm vir} = 0.5\,\rm{pc}$). In this case, the differences in number density profiles are more significant and result in appreciable differences in the eccentricity distribution of merging BBHs, as shown by Fig.\ref{fig:e_GC_rv}. Fig. \ref{fig:density_profile_rvcomp} also shows that although the higher ejection rate in the denser cluster reduces the difference in number density profiles with time,  significant differences remain even at $t=1\,\rm{Gyr}$. The central region of the $r_{\rm vir} = 0.5\,\rm{pc}$ model (with $r\lesssim 0.3\,\rm{pc}$, containing roughly half of all BHs of the $r_{\rm vir} = 0.5\,\rm{pc}$ cluster) maintains a considerably higher number density than any region in the  $r_{\rm vir} = 4\,\rm{pc}$ cluster model.

In Figure \ref{fig:density_profile_gc}, we vary the initial metallicity and keep the virial radius fixed, while in Figure \ref{fig:density_profile_rvcomp}, we keep the metallicity fixed and vary the virial radius. The differences in the radial profile of BH number density $n_\mathrm{BH}$ and the total number of BHs $N_\mathrm{BH}$ are far more pronounced when virial radius is changed, rather than metallicity. Further, the rate of decay of $N_\mathrm{BH}$ is far steeper for denser clusters. As can be seen in Fig. \ref{fig:e_GC_rv}, where we show eccentricity distributions of GW sources from three cluster models with identical initial conditions, different virial radii and therefore different $n_\mathrm{BH}$ have a significant impact on the dominance of different sub-channels of BBH merger formation in GCs. 

\bsp	
\label{lastpage}
\end{document}